\documentclass[11pt,a4paper,oneside]{article}
\usepackage[authoryear]{natbib}
\usepackage{times,epsfig,makeidx,amsfonts,amsmath,dcolumn,enumerate,multirow,graphicx,amssymb,colortbl,bm,url}
\usepackage[margin=1in]{geometry}
\usepackage{afterpage,lscape}
\usepackage{subfigure}
\usepackage{mathtools,upgreek}
\usepackage{paralist}

\newtheorem{definition}{Definition}

\begin{document}
\title{Survival and lifetime data analysis with a flexible class of distributions}
\author{{Francisco J.~Rubio\footnote{{\sc University of Warwick, Department of Statistics, Coventry, CV4 7AL, UK.} E-mail:  Francisco.Rubio@warwick.ac.uk}}\,  and Yili Hong\footnote{{\sc Virginia Tech, Department of Statistics, Blacksburg, VA 24061, USA.}}}

\maketitle

\begin{abstract}
We introduce a general class of continuous univariate distributions with positive support obtained by transforming the class of two-piece distributions. We show that this class of distributions is very flexible, easy to implement, and contains members that can capture different tail behaviours and shapes, producing also a variety of hazard functions. The proposed distributions represent a flexible alternative to the classical choices such as the log-normal, Gamma, and Weibull distributions. We investigate empirically the inferential properties of the proposed models through an extensive simulation study. We present some applications using real data in the contexts of time-to-event and accelerated failure time models. In the second kind of applications, we explore the use of these models in the estimation of the distribution of the individual remaining life.
\end{abstract}


\section{Introduction}\label{sec:intro}

In many areas, including medical applications, the quantities of interest take positive values. For instance, in survival analysis, the interest typically focuses on modelling the survival times of a group of patients in terms of a set of covariates (see \emph{e.g.}~\cite{L03}). Other areas where positive observations appear naturally are finance (\emph{e.g.}~in modelling the size of reinsurance claims), network traffic modelling \citep{M01}, reliability theory \citep{ME98}, environmental science \citep{MG10}, among many others. Parametric distributions provide a parsimonious way of describing the distribution of those quantities. Some of the most popular choices for modelling positive observations are the lognormal, log-logistic, Gamma, and Weibull distributions. We refer the readers to \cite{M12} for an extensive overview of these sorts of distributions as well as a study of their inferential properties in the presence of censored observations. However, these distributions do not always provide a good fit of the data. For example, when the data present heavier tails and/or a different shape around the mode than those captured by these distributions. In recent years, there has been an increasing interest in the development of flexible distributions with positive support in order to cover departures from the classical choices. Two popular strategies for generating new flexible distributions with positive support consist of:

\begin{inparaenum}[(i)]
\item \emph{Adding a shape parameter to an existing distribution with positive support}. For instance, in the context of reliability and survival analysis, \cite{MO97} proposed a transformation of a distribution $F(y;\theta)$, $y>0$, that introduces a new parameter $\gamma>0$. This transformation is defined through the cumulative distribution function (CDF)
\begin{eqnarray}\label{MOdistribution}
G(y;\theta,\gamma)=\dfrac{ F(y;\theta)}{F(y;\theta)+\gamma(1-F(y;\theta))}.
\end{eqnarray}

The interpretation of the parameter $\gamma$ is given in \cite{MO97} in terms of the behavior of the ratio of hazard rates of $F$ and $G$. This ratio is increasing in $y$ for $\gamma\geq 1$ and decreasing in $y$ for $0<\gamma\leq 1$. This transformation is then proposed for the exponential and Weibull distributions in \cite{MO97} in order to generate more flexible models for lifetime data. Clearly, for $\gamma=1$, $G$ and $F$ coincide. Many choices of $F(;\theta)$ have already been studied in the literature. We refer the reader to \cite{FS06} for a general mechanism for adding parameters to a distribution.

\item \emph{Using transformations from ${\mathbb R}$ to ${\mathbb R}_+$}. The most common choice for this transformation is the exponential function. The idea is to define a positive variable $Y$ by transforming a real variable $X$ through $Y=\exp(X)$. This method is used to produce the class of log-symmetric distributions. This is, the family of positive random variables such that their logarithm is symmetrically distributed. Some members of this class are the lognormal, log-logistic, and log-Student-$t$ distributions, which are obtained by transforming the normal, logistic, and Student-$t$ distributions, respectively. More recently, other families of distributions have been proposed by using this idea, such as the log Birnbaum-Saunders distribution \citep{B08}, log skew-elliptical distributions \citep{MG10}, log-generalised extreme value distributions \citep{RD15}, and log-scale mixtures of normals \citep{VS15}.

\end{inparaenum}
In this paper, we propose a new class of flexible distributions with support on ${\mathbb R}_+$ by applying the second method to the family of two-piece distributions \citep{FS98,A05,RS14}. In Section \ref{LTPDistributions}, we introduce the proposed class of distributions and show that it contains very flexible members that can capture a wide variety of shapes and tail behaviours. We show that these models can be seen as a subclass of composite models, which are of great interest in finance. The associated hazard functions are non-monotone with either increasing or decreasing right tails. These distributions are easy to implement using the R packages \citep{R13} `twopiece' and `TPSAS', which are available upon request. In Section \ref{MaximumLikelihoodEstimation}, we discuss the properties of the maximum likelihood estimators (MLEs) associated to these models. Although a formal study of the asymptotic properties of the proposed models is beyond the scope of this paper, we present a simulation study which reveals that adding a shape parameter, via two-piece transformations, has little effect on the performance of the maximum likelihood estimators. In Section \ref{NumericalExamples}, we present two kinds of applications with real data. In the first example, we illustrate the use of the proposed distributions in the context of data fitting. The main application is presented in the second and third examples, where we employ the proposed distributions for modelling the errors in an accelerated failure time model (AFT) with applications to medical data. In the third example, we discuss the use of a certain class of prediction intervals of the remaining life, which are informative for individual prognosis. In all of these examples, we discuss model selection between some appropriate competitors and the selection of the baseline distribution in the proposed family of distributions.

\section{Log Two-piece Distributions}\label{LTPDistributions}

For the sake of completeness, let us first recall the definition of two-piece distributions. Let $s(\cdot;\delta)$ be a symmetric unimodal density, with mode at $0$, with support on $\mathbb R$, and let $\delta\in\Delta\subset{\mathbb R}$ be a shape parameter (location and scale parameters can be added in the usual way). The corresponding CDF will be denoted as $S(\cdot;\delta)$. The shape parameter $\delta$ typically controls the tails of the density. For example, in the cases where $s(\cdot;\delta)$ is either a Student-$t$ density with $\delta>0$ degrees of freedom or an exponential power density with power parameter $\delta>0$ (see Appendix A).

\begin{definition}
A real random variable $X$ is said to be distributed according to a two-piece distribution if its probability density function (PDF) is given by (see \emph{e.g.}~\cite{RS14}):
\begin{eqnarray}\label{TP}
s_{tp}(x;\mu,\sigma_1,\sigma_2,\delta) = \dfrac{2}{\sigma_1+\sigma_2}\left[s\left(\dfrac{x-\mu}{\sigma_1};\delta\right) I(x<\mu) + s\left(\dfrac{x-\mu}{\sigma_2};\delta\right) I(x\geq \mu) \right].
\end{eqnarray}
\end{definition}
This is, a two-piece density is obtained by continuously joining two half-$s$ densities with different scale parameters on either side of the location $\mu$. The density (\ref{TP}) is unimodal, with mode at $\mu$, it is asymmetric for $\sigma_1\neq \sigma_2$, and coincides with the original density $s$ for $\sigma_1=\sigma_2$. Moreover, the tail behaviour of the PDF in (\ref{TP}) is the same in each direction, by construction. A popular reparameterisation is obtained by redefining $\sigma_1=\sigma a(\gamma)$ and $\sigma_2 = \sigma b(\gamma)$, where $a(\cdot)$ and $b(\cdot)$ are positive functions of the parameter $\gamma$ \citep{A05}. Two common choices for $a(\cdot)$ and $b(\cdot)$ are the inverse scale factors $\{a(\gamma),b(\gamma)\}=\{\gamma,1/\gamma\}$, $\gamma \in {\mathbb R}_+$ \citep{FS98}, and the epsilon-skew parameterisation $\{a(\gamma),b(\gamma)\}=\{1-\gamma,1+\gamma\}$, $\gamma \in (-1,1)$ \citep{MH00}. Other parameterisations are explored in \cite{RS14}. The PDF associated to this reparameterisation is given by
\begin{eqnarray}\label{TP2}
s_{tp}(x;\mu,\sigma,\gamma,\delta) = \dfrac{2}{\sigma[a(\gamma)+b(\gamma)]}\left[s\left(\dfrac{x-\mu}{\sigma b(\gamma)};\delta\right) I(x<\mu) + s\left(\dfrac{x-\mu}{\sigma a(\gamma)};\delta\right) I(x\geq \mu) \right]
\end{eqnarray}
This transformation preserves the existence of moments and the ease of use of the original distribution $s$. The corresponding cumulative distribution function and quantile function can be easily obtained from this expression (see \cite{A05}). This class of distributions has been shown to have good inferential properties for regular choices of the baseline density $s$ \citep{A05,JA10}.

By applying method (ii), described in Section~\ref{sec:intro}, to the family of two-piece distributions, we can produce distributions with support on ${\mathbb R}_+$ as follows.
\begin{definition}
A positive random variable $Y$ is said to be distributed according to a log two-piece (LTP) distribution if its PDF is given by:
\begin{align}\label{LTPPDF}
s_l(y; \mu,\sigma,\gamma,\delta) = \dfrac{2}{y\sigma[a(\gamma)+b(\gamma)]} &\left[ s\left(\dfrac{\log(y)-\mu}{\sigma b(\gamma)};\delta\right)I(y<e^{\mu})\right.\nonumber\\ &\quad\quad+\left.s\left(\dfrac{\log(y)-\mu}{\sigma a(\gamma)};\delta\right)I(y\geq e^{\mu}) \right].
\end{align}
\end{definition}
Given that the class of two-piece distributions contains all the symmetric unimodal distributions with support on the real line, it follows that the class of LTP distributions contains the class of log-symmetric distributions as well as models such that the distribution of $\log Y$ is asymmetric. The LTP Laplace distribution, which is obtained by using a Laplace baseline density $s$ in (\ref{LTPPDF}), has been studied in \cite{K01}. However, other types of log two--piece distributions have not been studied to the best of our knowledge. The corresponding CDF is given by
\begin{eqnarray}\label{LTPCDF}
S_l(y; \mu,\sigma,\gamma,\delta) &=& \dfrac{2b(\gamma)}{a(\gamma)+b(\gamma)}  S\left(\dfrac{\log(y)-\mu}{\sigma b(\gamma)};\delta\right)I(y<e^{\mu}) \notag\\
&+&\left[\dfrac{b(\gamma)-a(\gamma)}{a(\gamma)+b(\gamma)} + \dfrac{2a(\gamma)}{a(\gamma)+b(\gamma)}S\left(\dfrac{\log(y)-\mu}{\sigma a(\gamma)};\delta\right)\right]I(y\geq e^{\mu}) .
\end{eqnarray}

We can observe that the ratio of the mass cumulated on either side of the value $y=e^{\mu}$ is given by
\begin{eqnarray*}
R(\gamma) &=& \dfrac{S_l(e^{\mu}; \mu,\sigma,\gamma,\delta)}{1-S_l(e^{\mu}; \mu,\sigma,\gamma,\delta)} = \dfrac{b(\gamma)}{a(\gamma)}.
\end{eqnarray*}

\begin{figure}[h]
\begin{center}
\begin{tabular}{c c}
\psfig{figure=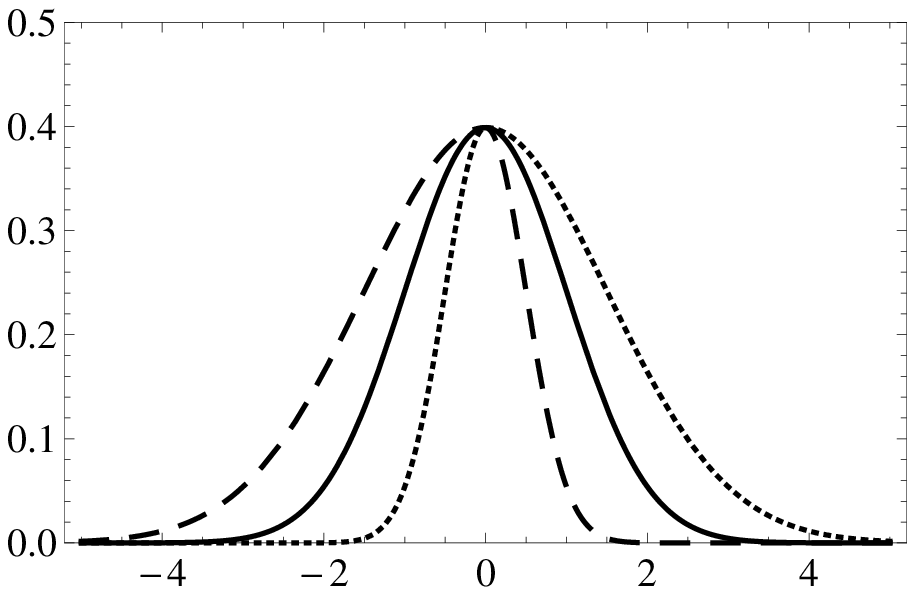,  height=5cm, width = 6cm}  &
\psfig{figure=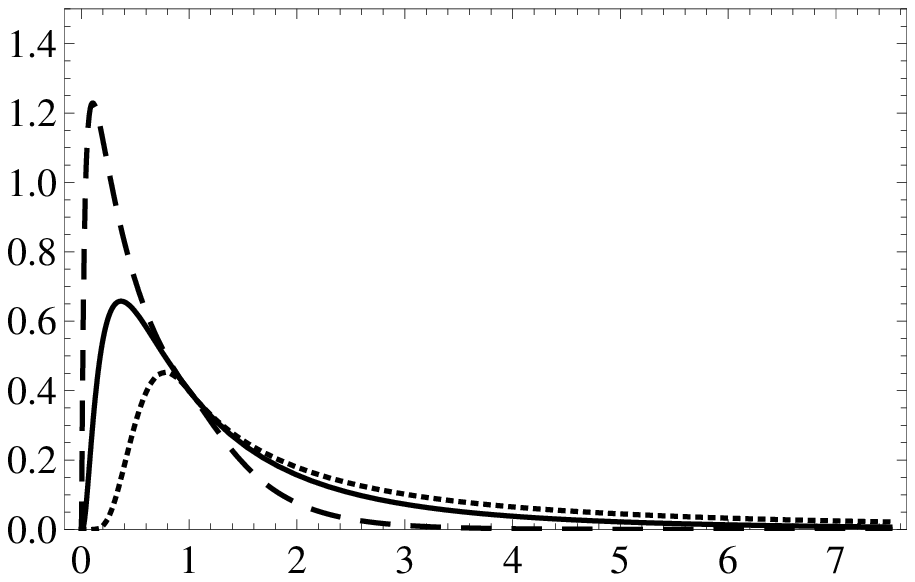,  height=5cm, width = 6cm}\\
(a) & (b)
\end{tabular}
\end{center}
\caption{\small (a) Two-piece normal densities (epsilon-skew parameterisation) $\mu=0$, $\sigma=1$, $\gamma=-0.5,0,0.5$, and (b) Log two-piece normal densities $\mu=0$, $\sigma=1$, $\gamma=-0.5,0,0.5$).}
\label{fig:LTPN}
\end{figure}

\begin{figure}[h]
\begin{center}
\begin{tabular}{c c}
\psfig{figure=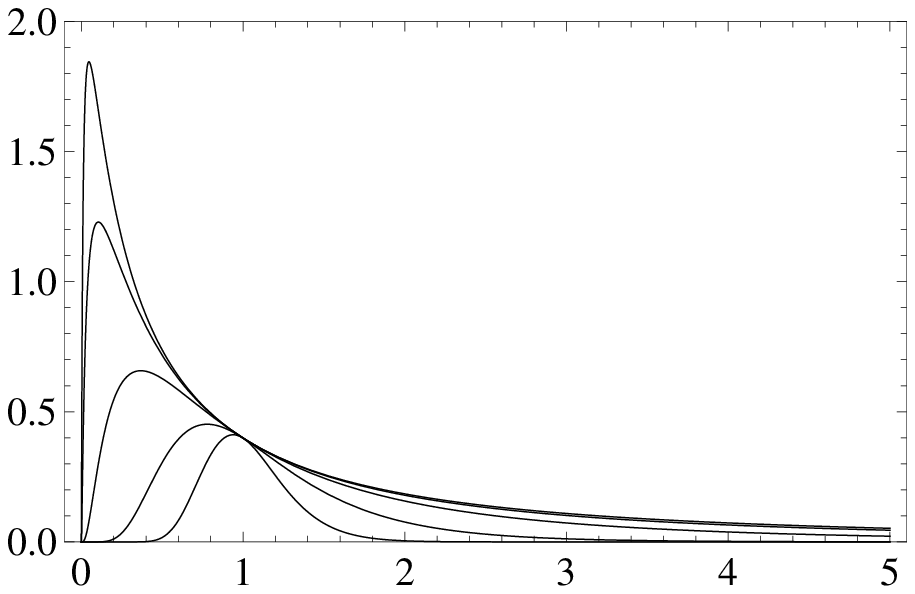,  height=5cm, width = 6cm}  &
\psfig{figure=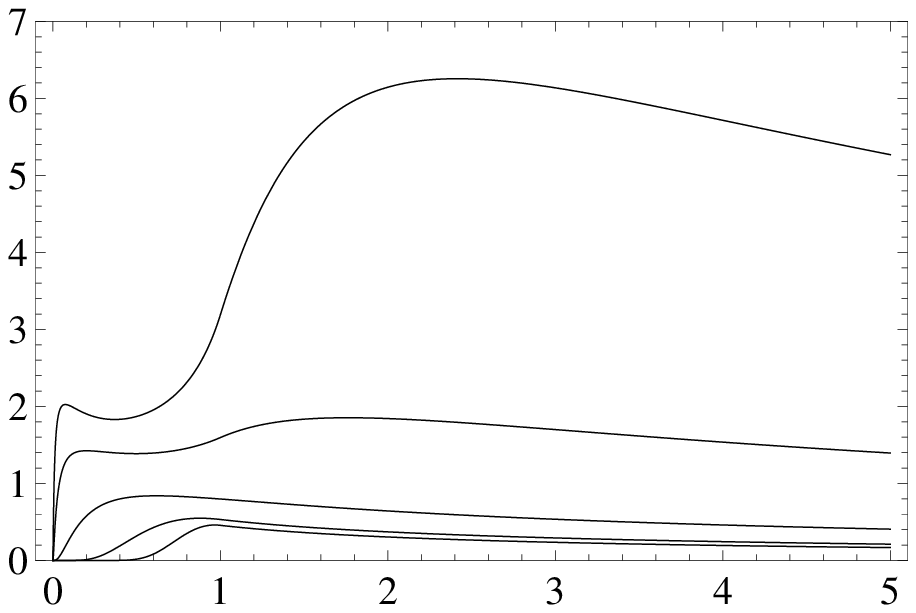,  height=5cm, width = 6cm}\\
(a) & (b)\\
\psfig{figure=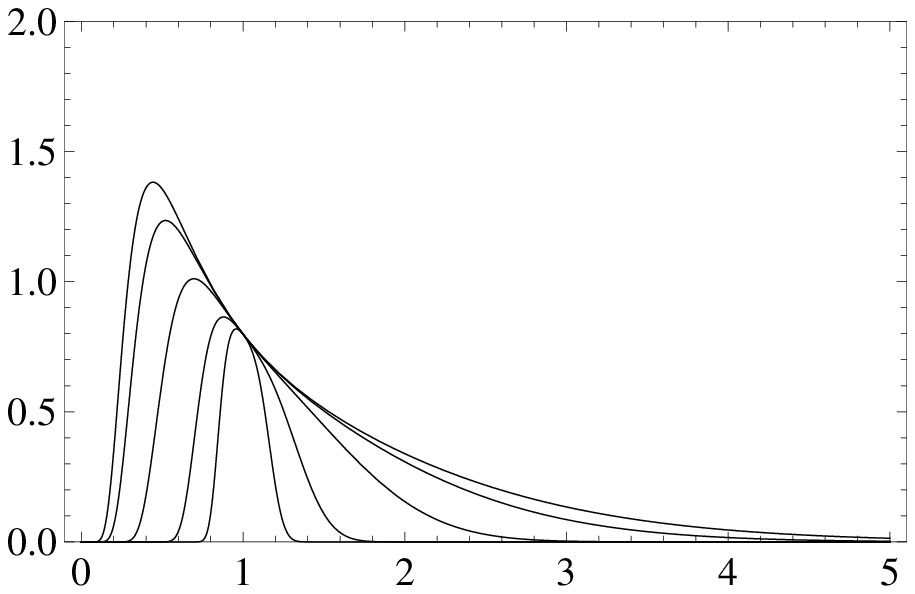,  height=5cm, width = 6cm}  &
\psfig{figure=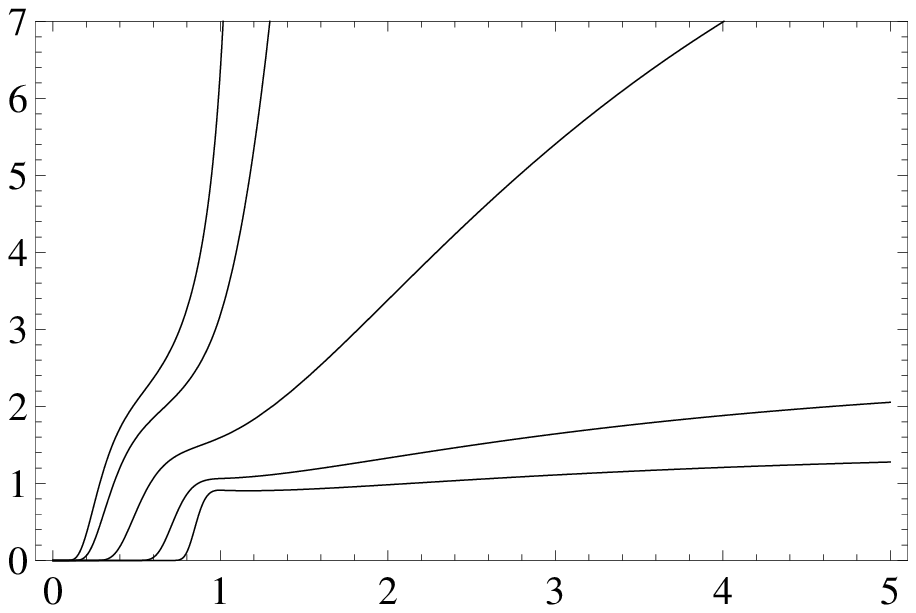,  height=5cm, width = 6cm}\\
(c) & (d)\\
\psfig{figure=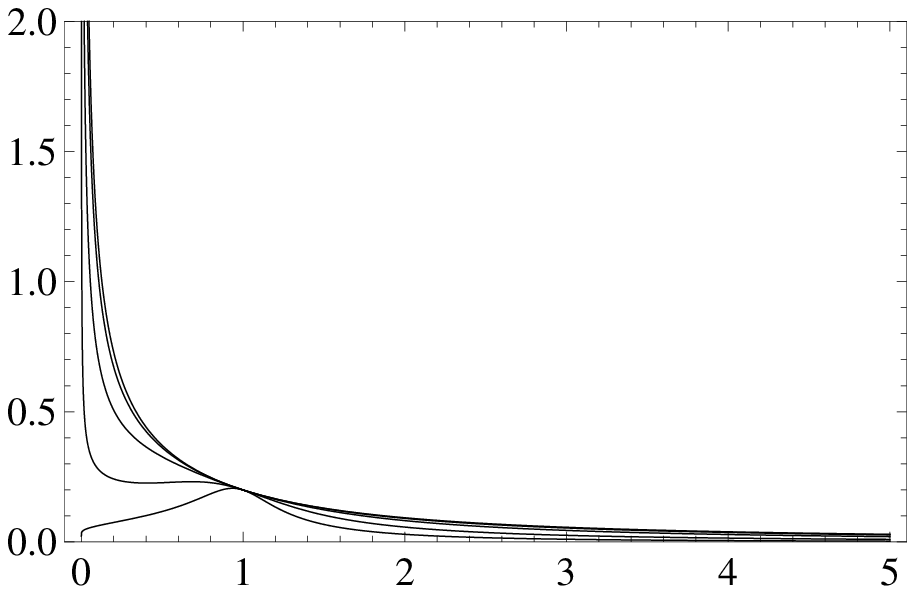,  height=5cm, width = 6cm}  &
\psfig{figure=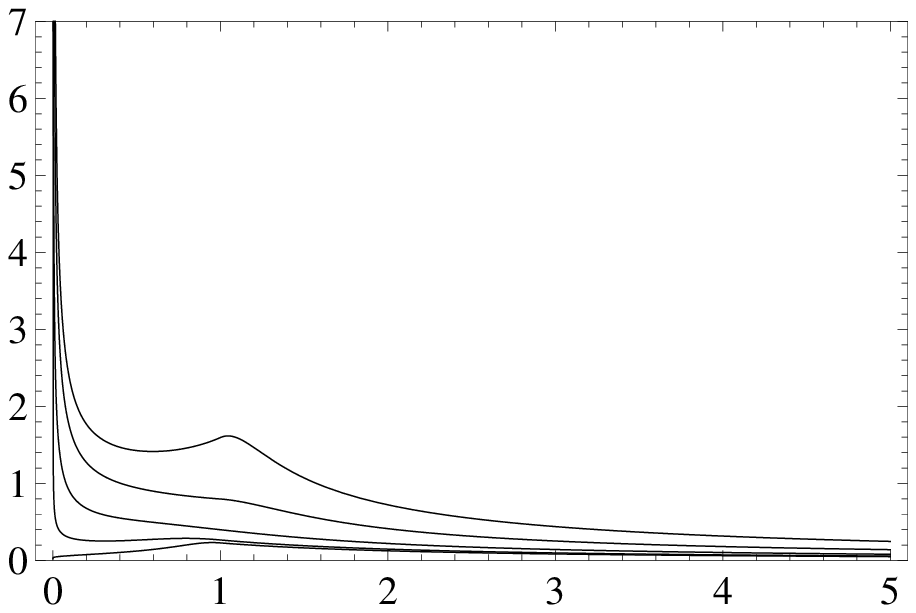,  height=5cm, width = 6cm}\\
(e) & (f)
\end{tabular}
\end{center}
\caption{\small Examples of LTP densities (left side) and hazard functions (right side) ($\mu=0$, $\sigma=1$, $\gamma=-0.75,-0.5,0,0.5,0.75$). (a,b) $\delta=1$; (c,d) $\delta=2$; (e,f) $\delta=0.5$.}
\label{fig:DH}
\end{figure}

This helps us to identify the different roles of the parameters $\gamma$ and $\delta$. The parameter $\gamma$ controls the cumulation of mass on either side of $y=e^{\mu}$, while the parameter $\delta$ controls the tails of the density. In Figure \ref{fig:LTPN}a we present some examples of a two-piece normal PDF with different values of the parameter $\gamma$. In these cases, the parameter $\gamma$ only affects the asymmetry of the density. Figure \ref{fig:LTPN}b shows the corresponding LTP normal PDFs. We can observe that in these cases the parameter $\gamma$ affects the shapes of the density. That is, it controls the mass cumulated above and below the value $y=1$ as well as the spread and mode of the density. The corresponding hazard function can be easily constructed from (\ref{LTPPDF}) and (\ref{LTPCDF}). Figure \ref{fig:DH} shows the variety of shapes of the density and hazard functions obtained for a log two-piece sinh-arcsinh distribution (LTP SAS, which is obtained by using a symmetric sinh-arcsinh baseline density function in (\ref{TP2}), see also \cite{R15}. The corresponding expression is provided in Appendix A). The implementation of LTP distributions is straightforward in R by using the packages `twopiece' and `TPSAS', which are freely available upon request. Moreover, the $p$th moment of a LTP distribution exists, whenever the $p$th moment of the underlying (log-symmetric) log-$s$ distribution exists. In particular, all moments of the LTP normal distribution exist. 

\subsection*{An alternative construction}

The family of two-piece distributions (\ref{TP})-(\ref{TP2}) can be seen as a special kind of finite mixtures of truncated PDFs, as shown in \citep{RS14}. In a similar fashion, the family of log two-piece distributions can be obtained as a particular class of finite mixtures of truncated distributions with positive support. In the context of survival and size distributions these sorts of mixtures are known as composite models (see \cite{NB13} for a literature review). Recall first that the PDF of a composite model can be written as:
\begin{eqnarray}\label{CompPDF}
s_c(y) = \omega s_1^*(y) I(y\leq \theta) + (1-\omega) s_2^*(y) I(y>\theta),\,\,\, y>0,
\end{eqnarray}
where $\omega = \dfrac{s_2(\theta)S_1(\theta)}{s_2(\theta)S_1(\theta) + s_1(\theta)[1-S_2(\theta)]}$, $\theta>0$ is a threshold parameter, $s_1^*(y)=\dfrac{s_1(y)}{S_1(\theta)}$, $s_2^*(y)=\dfrac{s_2(y)}{1-S_2(\theta)}$, $s_1$ and $s_2$ are continuous PDFs with support on ${\mathbb R}_+$, and $S_1$ and $S_2$ are the corresponding CDFs. If we fix $s_1(y)=\dfrac{1}{\sigma_1 y}s\left(\dfrac{\log(y)-\mu}{\sigma_1}\right)$, $s_2(y)=\dfrac{1}{\sigma_1 y}s\left(\dfrac{\log(y)-\mu}{\sigma_2}\right)$, for some symmetric density $s$ with support on ${\mathbb R}$, and $\theta=e^{\mu}$, then it follows that (\ref{CompPDF}) coincides with (\ref{LTPPDF}), up to a reparameterisation.

From this alternative construction, we conclude that the family of LTP distributions represents a subclass of composite models with the appealing properties and interpretability of parameters discussed above. This also allows us to motivate the use of LTP distributions as survival and size distributions.

\section{Models and Maximum Likelihood Estimation}\label{MaximumLikelihoodEstimation}
In this section, we present the parameter estimation procedure for time-to-event and accelerated failure time (AFT) models.

\subsection{Time-to-event model}

Let ${\bf T}=(T_1,\dots,T_n)$ be an independent sample of survival times distributed as in (\ref{LTPPDF}). The likelihood function of the parameters $(\mu,\sigma,\gamma,\delta)$ is defined as:
\begin{eqnarray*}
L(\mu,\sigma,\gamma,\delta) = \prod_{j=1}^n s_l(T_j;\mu,\sigma,\gamma,\delta).
\end{eqnarray*}
The MLE is defined as the parameter values that maximise the likelihood function. By noting that
\begin{eqnarray*}
L(\mu,\sigma,\gamma,\delta) \propto \prod_{j=1}^n s_{tp}(\log(T_j);\mu,\sigma,\gamma,\delta),
\end{eqnarray*}
it follows that the MLEs of the parameters of LTP distributions are the same as the MLEs of the parameters of the underlying two-piece distribution for the sample $\log({\bf T}) = [\log(T_1),\dots,$ $\log(T_n)]$. Inferential aspects of 3- and 4-parameter two-piece distributions have been largely discussed. For example, \cite{A05} show that, under certain regularity conditions on the baseline density $s$ in (\ref{TP2}), the maximum likelihood estimators of the parameters of these distributions are consistent and asymptotically normal under the epsilon-skew parameterisation. \cite{JA10} and \cite{RS14} study some parameterisations that induce parameter orthogonality between the parameters $\mu$ and $\sigma$, showing that the epsilon-skew parameterisation induces this property. Parameter orthogonality, in turn implies a good asymptotic behaviour of the MLE \citep{JA10}. In most cases, the MLE is not available in closed-form, and it has to be obtained numerically.

Samples containing censored observations are common in the context of survival analysis. The most common types of censoring in this context correspond to:
\begin{enumerate}[(i)]
\item Left-censoring: when the phenomenon of interest has already occurred before the start of the study. A left-censored observation is an interval of the type $[0,T_j)$, where $T_j$ represents the start of the study for subject $j$.

\item Interval censoring: when the phenomenon of interest occurs within a finite period of time $[T_j^L,T_j^R]$.

\item Right-censoring: when the phenomenon of interest is not observed during the duration of the study. A right-censored observation is an interval of the type $(T_j,\infty]$, where $T_j$ represents the duration of the study for subject $j$.
\end{enumerate}

Ignoring censoring induces bias in the estimation of the parameters. Different types of censoring imply different contributions of the observations to the likelihood function. The contribution of a left-censored observation to the likelihood is $S_l(T_j;\mu,\sigma,\gamma,\delta)$; while the contribution of an interval-censored observation to the likelihood is $S_l(T_j^R;\mu,\sigma,\gamma,\delta)-S_l(T_j^L;\mu,\sigma,\gamma,\delta)$; and the contribution of a right-censored observation to the likelihood is $1-S_l(T_j;\mu,\sigma,\gamma,\delta)$. If we define the sets $\text{Left}=\{j: T_j \text{ is left-censored}\}$, $\text{Int}=\{j: T_j \text{ is interval-censored}\}$, $\text{Right}=\{j: T_j \text{ is right-censored}\}$, and $\text{Obs}=\{j: T_j \text{ in uncensored}\}$, then we can write the likelihood function as follows:
\begin{align*}
L(\mu,\sigma,\gamma,\delta) = &\prod_{j\in\text{Obs}} s_l(T_j;\mu,\sigma,\gamma,\delta) \times \prod_{j\in\text{Left}} S_l(T_j;\mu,\sigma,\gamma,\delta) \\&\times  \prod_{j\in\text{Right}}\left[1-S_l(T_j;\mu,\sigma,\gamma,\delta)\right]
\times  \prod_{j\in\text{Int}} \left[ S_l(T_j^R;\mu,\sigma,\gamma,\delta)-S_l(T_j^L;\mu,\sigma,\gamma,\delta)\right].
\end{align*}

The latter expression emphasises the practical importance of using distributions with a tractable distribution function.

\subsection{Accelerated failure time models}

AFT models are a useful tool for modelling the set of survival times ${\bf T}=(T_1,\dots,T_n)$ in terms of a set of covariates $\bm{\beta} = (\beta_1,\dots,\beta_p)$ through the model equation:
\begin{eqnarray}\label{AFTModel}
h(T_j) = x_j^{\top}\bm{\beta} + \varepsilon_j, \,\,\, j=1,\dots,n,
\end{eqnarray}

\noindent where ${\bf X}= (x_1,\dots,x_n)^{\top}$ is an $n\times p$ known design matrix and $\varepsilon_j \stackrel{ind.}{\sim} F(\cdot;\bm{\theta})$, $F$ is a continuous distribution with support on ${\mathbb R}$ and parameters $\bm{\theta}\in \Theta \subset {\mathbb R}^d$, and $h:{\mathbb R}_+\rightarrow{\mathbb R}$ is a continuous increasing function. The most common choice for $h$ is the logarithmic function, while the distribution of the errors $\varepsilon_j$ is typically assumed to be normal. Given that the assumption of normality of the errors can be restrictive in practice, other distributional assumptions have been recently studied such as the log Birnbaum-Saunders \citep{B08}, finite mixtures of normal distributions \citep{KL08}, the symmetric family of scale mixtures of normals \citep{VS15}, and the log-generalised extreme value distribution \citep{RD15}.

AFT models are extremely relevant in medicine, given that survival data naturally arise in many medical studies, which typically involve the follow-up of other covariates. The presence of different types of censored observations is common in this context \citep{B08,KL08,VS15}. If we assume that the errors $\varepsilon_j$ are distributed according to a LTP distribution with location $0$ and $\bm{\theta}=(\sigma,\gamma,\delta)$, then we can write the likelihood function as follows,
\begin{align*}
L(\bm{\beta},\sigma,\gamma,\delta) =& \prod_{j\in\text{Obs}} s_l(T_j;x_j^{\top}\bm{\beta},\sigma,\gamma,\delta) \times \prod_{j\in\text{Left}} S_l(T_j;x_j^{\top}\bm{\beta},\sigma,\gamma,\delta)\\ &\times  \prod_{j\in\text{Right}}\left[1-S_l(T_j;x_j^{\top}\bm{\beta},\sigma,\gamma,\delta)\right]\\
&\times  \prod_{j\in\text{Int}} \left[ S_l(T_j^R;x_j^{\top}\bm{\beta},\sigma,\gamma,\delta)-S_l(T_j^L;x_j^{\top}\bm{\beta},\sigma,\gamma,\delta)\right],
\end{align*}

\noindent with the notation discussed previously. It is important to notice that by using asymmetric errors, we obtain a curve that does not represent the mean response. However, as discussed in \cite{AG08}, this lack of centring can be calibrated after estimating the parameters by adding a suitable quantity $M_{\varepsilon}$ which reflects the lack of centring of the errors. For instance, in order to obtain the mean response, we can use $M_{\varepsilon}=-{\mathbb E}[\varepsilon_j]$, computed at the MLE of the parameters of the error distribution. This strategy will only affect the intercept parameter. When using baseline models with infinite variance (such as a log-Cauchy distribution), one might opt for centring around the median (or another quantile), instead of the mean. A formal study of the asymptotic properties of the MLEs under different types of censoring is beyond the scope of this paper. However, in Section \ref{SimulationStudy}, we illustrate the performance of the MLEs in a linear regression model with censored observations through a simulation study.

\section{Simulation Study}\label{SimulationStudy}

In this section, we present a simulation study in order to illustrate the performance of the MLEs of the parameters of some LTP distributions. Throughout, we employ the epsilon-skew parameterisation discussed previously. In our first simulation scenario, we simulate $N=10,000$ samples of sizes $n=30,50,100,250,500,1000$ from a LTP normal (log two-piece normal) with different combinations of the parameter values: $\mu=0$, $\sigma=1$, and $\gamma=0,0.25,0.5,0.75$. Negative values of $\gamma$ would induce similar results, since they produce the corresponding reflected density about $e^{\mu}$, and are therefore omitted. For each of these samples, we calculate the corresponding MLEs, using the R command `optim', and calculate the bias, variance, and root-mean-square error (RMSE) of these. In our second simulation scenario, we simulate $N=10,000$ samples from a LTP $t$ (log two-piece Student-$t$) with parameters: $\mu=0$, $\sigma=1$, $\gamma=0,0.25,0.5,0.75$, and $\delta=1$. The third simulation scenario is analogous to the second scenario, with $\delta=2$. In the fourth scenario, we simulate $N=10,000$ samples from a LTP SAS with parameters : $\mu=0$, $\sigma=1$, $\gamma=0,0.25,0.5,0.75$, and $\delta=0.75$. Tables \ref{table:LTPN}--\ref{table:LTPSAS2} present the results of these simulations.

In second class of simulations, we investigate the performance of the use of log two-piece errors in AFT models (\ref{AFTModel}). For this purpose, we simulate from the linear regression model:
\begin{eqnarray*}
\log(y_j) = {\bf x}_j^{\top}\bm{\beta} + \varepsilon_j, \,\,\, j=1,\dots,n,
\end{eqnarray*}
with $n=100,250,500$, $\beta = (1,2,3)^{\top}$, and ${\bf x}_j = (1,x_{j1},x_{j2})^{\top}$. The second and third entries of the covariates ${\bf x}_j$ are simulated from a right-half-normal with scale parameter $1/3$. For the distribution of the errors $\varepsilon_j$ we consider the following cases: (i) a two-piece normal distribution with parameters $\mu=0$, $\sigma=0.25$, and $\gamma=0,0.25,0.5$; and (ii) a TP SAS distribution \citep{R15} with parameters $\mu=0$, $\sigma=0.25$, $\gamma=0,0.25,0.5$, and $\delta=0.75$. We truncate the observations $y_j$ that are greater than $17.5$. This censoring mechanism produces samples with 15\%--35\% censored observations. Tables \ref{table:LTPNR}--\ref{table:LTPSASR} present the results of these simulations.

The overall conclusions of this extensive simulation study are that the value of the shape parameter $\gamma$ does not seem to greatly affect the performance of the MLEs, while the use of models with a tail parameter $\delta$ have a clear effect on the performance of the MLEs.  The performance of the MLEs of $\delta$ in LTP $t$ and LTP SAS models for small samples is different: the bias is smaller in the LTP SAS model. However, the estimation of $\sigma$ is more accurate in the LTP $t$ model. This is, perhaps, an unsurprising conclusion, given that it is well--known that it is difficult to learn about tail parameters with small samples and that tail parameters control the tail behaviour differently in different models. However, this analysis helps us to quantify the order of observations required for an accurate estimation. For LTP models with 4 parameters, such as the LTP $t$ and LTP SAS models, it is necessary to have at least $200$ observations in order to accurately estimate the tail parameters. In fact, the proposed flexible models are not recommended with small samples since, intuitively, these do not contain information about the features captured by the shape parameters $\gamma$ and $\delta$.
\clearpage
\afterpage{
\begin{landscape}
\begin{table}[!htbp]
\begin{center}
\caption{\small Simulation results: LTP Normal.}
\label{table:LTPN}
\vspace{1em}
\begin{tabular}[h]{|c|c|c|c|c|c|c|c|c|c|c|c|c|c|}
\hline
Par. & & \multicolumn{3}{c|}{$\mu=0$, $\sigma=1$, $\gamma=0$} & \multicolumn{3}{c|}{$\mu=0$, $\sigma=1$, $\gamma=0.25$} & \multicolumn{3}{c|}{$\mu=0$, $\sigma=1$, $\gamma=0.5$} & \multicolumn{3}{c|}{$\mu=0$, $\sigma=1$, $\gamma=0.75$}\\
\hline
 & $n$ & $\hat\mu$ & $\hat\sigma$ & $\hat\gamma$ & $\hat\mu$ & $\hat\sigma$ & $\hat\gamma$ & $\hat\mu$ & $\hat\sigma$ & $\hat\gamma$ & $\hat\mu$ & $\hat\sigma$ & $\hat\gamma$\\
\hline
\multirow{6}{*}{Bias} &  30 & -0.0017 & 0.0483 & -0.0006 & -0.0277 & 0.0498 & -0.0329& -0.0436 & 0.0536 & -0.0616 & -0.0296 & 0.0570 & -0.0734\\
& 50 &  0.0045 & 0.0277 &  0.0018 & -0.0277  & 0.0283 & -0.0244 & -0.0431 &  0.0304 & -0.0430 & -0.0355 & 0.0342 & -0.0517\\
& 100 &  -0.0018 & 0.0127 &  -0.0007  & -0.0148 & 0.0134 & -0.0111 & 0.0270 & 0.0140 & -0.0219 & -0.0306 &0.0158  &  -0.0303\\
& 250 & -0.0001  & 0.0045 &  -0.0003 & -0.0039  & 0.0046 &  -0.0037 & -0.0079  & 0.0050 &  -0.0071 & -0.0103  & 0.0052 & -0.0100\\
& 500 & 0.0006  & 0.0025 & 0.0002 & -0.0013  & 0.0025 &  -0.0015 & -0.0032  & 0.0028 & -0.0033 & -0.0046  & 0.0025 & -0.0047 \\
& 1000 & 0.0003  & 0.0013 & 0.0002  &  0.0001 & 0.0014 &  -0.0001 &  -0.0006 & 0.0013 & -0.0008 &  -0.0015 & 0.0012 & -0.0018 \\
\hline
\multirow{6}{*}{Variance} &  30 & 0.3824 & 0.0171 & 0.1588 & 0.3576 & 0.0172 &  0.1495 & 0.2760 & 0.0170 &  0.1151  & 0.1357 & 0.0166 & 0.0565\\
& 50 &  0.2023 & 0.0098 &  0.0759 & 0.1865 & 0.0098 & 0.0709 & 0.1499 & 0.0099 & 0.0580 & 0.0839 & 0.0099 & 0.0331\\
& 100 &  0.0786 & 0.0048 &  0.0277 & 0.0772 & 0.0048 & 0.0273 & 0.0673 & 0.0049 & 0.0240 & 0.0441 & 0.0050 & 0.0162\\
& 250 &  0.0279 & 0.0020 & 0.0095  & 0.02640  & 0.0020 & 0.0089 &  0.0224 & 0.0020 & 0.0076 & 0.0148  & 0.0019 &  0.0050 \\
& 500 & 0.0138  & 0.0010 &  0.0046 & 0.0128  & 0.0009 &  0.0043  &  0.0107 & 0.0010 &  0.0036 & 0.0066  & 0.0010 & 0.0022\\
& 1000 & 0.0067  &  0.0005 & 0.0022 &  0.0064 & 0.0005 &  0.0021 &  0.0052 &  0.0004 & 0.0017 &  0.0031 & 0.0005 & 0.0010 \\
\hline
\multirow{6}{*}{RMSE} &  30 & 0.6184 & 0.1397 & 0.3985 & 0.5986 & 0.1403 & 0.3881 & 0.5271 & 0.1411 & 0.3449 & 0.3696 & 0.1411 & 0.2488\\
& 50 & 0.4498  & 0.1030 &  0.2756 & 0.4327 & 0.1033 & 0.2674 & 0.3896  & 0.1044 & 0.2447 & 0.2918 & 0.1056 & 0.1892\\
& 100 & 0.2805  & 0.0711 & 0.1665 & 0.2782 & 0.0711 & 0.1656 & 0.2609 & 0.0718 & 0.1567  & 0.2122 & 0.0728 & 0.1309 \\
& 250 &  0.1671 & 0.0450 & 0.0977 &  0.1625 & 0.0452 & 0.0949 &  0.1500 & 0.0452 &  0.0875 & 0.1224  & 0.0448 &  0.0717 \\
& 500 & 0.1178  & 0.0317 & 0.0682  & 0.1135  & 0.0315 &  0.0659 & 0.1037  & 0.0319 & 0.0602 &  0.0817  & 0.0320 & 0.0474 \\
& 1000 & 0.0821  & 0.0222 & 0.0472 &  0.0800 & 0.0222 &  0.0462 &  0.0721 & 0.0222 & 0.0415 & 0.0559  & 0.0223 & 0.0324 \\
\hline
\end{tabular}
\end{center}
\end{table}
\end{landscape}
}

\clearpage

\begin{table}[!htbp]
\begin{center}
\caption{\small Simulation results: LTP $t$, $\delta=1$.}\label{table:LTPt11}
\vspace{.5em}
{\small
\begin{tabular}[h]{|c|c|c|c|c|c|c|c|c|c|}
\hline
Par. & & \multicolumn{4}{c|}{$\mu=0$, $\sigma=1$, $\gamma=0$, $\delta=1$} & \multicolumn{4}{c|}{$\mu=0$, $\sigma=1$, $\gamma=0.25$, $\delta=1$} \\
\hline
 & $n$ & $\hat\mu$ & $\hat\sigma$ & $\hat\gamma$ & $\hat\delta$ & $\hat\mu$ & $\hat\sigma$ & $\hat\gamma$ & $\hat\delta$ \\
\hline
\multirow{6}{*}{Bias} &  30 & -0.0012 & -0.0358 & 0.0005 & -197.7 & -0.0149  & -0.0340  & -0.0118  & -290.6 \\
& 50  & 0.0083 & -0.0237 & 0.0032 & -1.18 & 0.0018  &  -0.0218 & -0.0024  & -4.8 \\
& 100 & -0.0033 & -0.0082 & -0.0019 & -0.0497 &  -0.0040 & -0.0097  & -0.0031  & -0.0497 \\
& 250 & 0.0010 & -0.0004 & 0.0001 & -0.0155 & 0.0013  & -0.0010  & -0.0005  & -0.0163 \\
& 500 & 0.0015 & -0.0008 & 0.0010 & -0.0085 & 0.0005  &  -0.0002 & 0.0003  &  -0.0081\\
& 1000 & 0.0004 & -5$\times10^{-5}$ & 0.0003 & -0.0037 & 0.0004  &  0.0003 &  -0.0001 & -0.0038 \\
\hline
\multirow{6}{*}{Var.} &  30 & 0.2564 & 0.1322 & 0.0703 & 9$\times10^7$ &  0.2452 & 0.1343  &  0.0678 & 1$\times10^8$ \\
& 50 & 0.1220 & 0.0701 & 0.0362 & 1$\times10^4$ &  0.1110 &  0.0688 & 0.0335  & 9$\times10^4$ \\
& 100 & 0.0501 & 0.0304 & 0.0164 & 0.0467 & 0.0464  & 0.0309  &  0.0152 & 0.0487 \\
& 250 & 0.0179 & 0.0114 &  0.0059 & 0.0142 & 0.0170  &  0.0113 &  0.0055 & 0.0142 \\
& 500 & 0.0090 & 0.0056 & 0.0030 & 0.0065 & 0.0084  &  0.0056 & 0.0028  & 0.0065 \\
& 1000& 0.0044 & 0.0028 &  0.0014 & 0.0032 &  0.0041 &  0.0028  &  0.0013 & 0.0032 \\
\hline
\multirow{6}{*}{RMSE} &  30  & 0.5064 & 0.3653 & 0.2651 & 9510.9 & 0.4954  & 0.3680  & 0.2607  & 1$\times10^4$ \\
& 50 & 0.3494 & 0.2658 & 0.1903 & 103.7 & 0.3331  &  0.2632 & 0.1831  & 305.8 \\
& 100 & 0.2238 & 0.1745 & 0.1281 & 0.2217 & 0.2155  &  0.1760 & 0.1234  & 0.2263 \\
& 250 & 0.1340 & 0.1068 & 0.0769 & 0.1202 & 0.1307  & 0.1063  & 0.0743  & 0.1204 \\
& 500 & 0.0948 & 0.0754 & 0.0549 & 0.0810 &  0.0917 & 0.0753  &  0.0529 & 0.0811 \\
& 1000 & 0.0669 & 0.0530 & 0.0383 & 0.0571 & 0.0646  &  0.0529 &  0.0369 & 0.0568 \\
\hline
\end{tabular}
}
\end{center}
\end{table}

\begin{table}[!htbp]
\begin{center}
\caption{\small Simulation results: LTP $t$, $\delta=1$.}
\label{table:LTPt12}
\vspace{.5em}
{\small
\begin{tabular}[h]{|c|c|c|c|c|c|c|c|c|c|}
\hline
Par. & & \multicolumn{4}{c|}{$\mu=0$, $\sigma=1$, $\gamma=0.5$, $\delta=1$} & \multicolumn{4}{c|}{$\mu=0$, $\sigma=1$, $\gamma=0.75$, $\delta=1$} \\
\hline
 & $n$ & $\hat\mu$ & $\hat\sigma$ & $\hat\gamma$ & $\hat\delta$ & $\hat\mu$ & $\hat\sigma$ & $\hat\gamma$ & $\hat\delta$ \\
\hline
\multirow{6}{*}{Bias} &  30 & -0.0186 & -0.0284 & -0.0239 & -134.3 & -0.0002  & -0.0125  &  -0.0332 & -60.5 \\
& 50  & -0.0040 & -0.0184 & -0.0091 & -0.2530 & -0.0030  & -0.0116  &  -0.0175 &  -0.1832\\
& 100 & -0.0036 & -0.0110 & -0.0046 & -0.0513 &  0.0012 &  -0.0087  &  -0.0065 & -0.0490 \\
& 250 & 0.0008 & -0.0016 & -0.0012 & -0.0165 &  0.0005 &  -0.0015 & -0.0028  & -0.0173 \\
& 500 & 0.0008 & 0.0002 & -3$\times10^{-5}$ & -0.0080 &  0.0004 &  0.0011 &  -0.0011 & -0.0073 \\
& 1000 & 0.0003 & 0.0005 &-0.0003  & -0.0037 &  0.0004 &  0.0001 & -0.0006  &  -0.0039\\
\hline
\multirow{6}{*}{Var.} &  30 & 0.1953 & 0.1331 & 0.0536 & 7$\times10^7$ & 0.1089 & 0.1255 & 0.0304  & 3$\times10^6$ \\
& 50 & 0.0975 & 0.0683 & 0.0279 & 70.8 &   0.0615 &  0.0684 &  0.0171 & 30.1  \\
& 100 & 0.0394 & 0.0313 & 0.0124 & 0.0628 & 0.0268  & 0.0314  &  0.0078 & 0.0498 \\
& 250 & 0.0140 & 0.0113 & 0.0044 & 0.0144 &  0.0093 & 0.0114  & 0.0028  & 0.0145 \\
& 500 & 0.0068 & 0.0056 &  0.0022 & 0.0064 & 0.0041  & 0.0055  & 0.0013  & 0.0065 \\
& 1000& 0.0033 & 0.0028 & 0.0011 & 0.0031 & 0.0020  &   0.0028 & 0.0006  & 0.0031 \\
\hline
\multirow{6}{*}{RMSE} &  30  & 0.4423 & 0.3659 &  0.2327 & 8814.4 & 0.3299  & 0.3544  & 0.1777  & 1870.2 \\
& 50 & 0.3123 & 0.2621 & 0.1674 & 8.4 &  0.2481 & 0.2619  &  0.1319 & 5.5 \\
& 100 & 0.1985 & 0.1774 & 0.1113 & 0.2558 &  0.1637 & 0.1773  & 0.0885  & 0.2285 \\
& 250 & 0.1183 & 0.1063 & 0.0667 & 0.1211 & 0.0964  &  0.1068  & 0.0534  & 0.1216 \\
& 500 & 0.0828 & 0.0751 & 0.0477 & 0.0809 &  0.0645 & 0.0746  & 0.0362  & 0.0810 \\
& 1000 & 0.0581 & 0.0530 & 0.0331 & 0.0565 & 0.0446  &  0.0530 & 0.0253  & 0.0563 \\
\hline
\end{tabular}
}

\end{center}
\end{table}

\begin{table}[!htbp]
\begin{center}
\caption{\small Simulation results: LTP $t$, $\delta=2$.}\label{table:LTPt21}
{\small
\begin{tabular}[h]{|c|c|c|c|c|c|c|c|c|c|}
\hline
Par. & & \multicolumn{4}{c|}{$\mu=0$, $\sigma=1$, $\gamma=0$, $\delta=2$} & \multicolumn{4}{c|}{$\mu=0$, $\sigma=1$, $\gamma=0.25$, $\delta=2$} \\
\hline
 & $n$ & $\hat\mu$ & $\hat\sigma$ & $\hat\gamma$ & $\hat\delta$ & $\hat\mu$ & $\hat\sigma$ & $\hat\gamma$ & $\hat\delta$ \\
\hline
\multirow{6}{*}{Bias} &  30 & 0.0052 & -0.0347 & 0.0018 & -2858.8 &  -0.0205 &  -0.0350 & -0.0197  & -2502.9 \\
& 50  & -0.0026 & -0.0269 & -0.0017 & -623.9 &  -0.0168 & -0.0247  &  -0.0114 & -476.8 \\
& 100 & 0.0005 & -0.0096 & 0.0005 & -5.8 & -0.0047  &  -0.0090 &  -0.0037 & 27.0 \\
& 250 & 0.0010 & -0.0028 & -0.0001 & -0.0883 &  0.0008 &  -0.0035 &  -0.0007 & -0.0890 \\
& 500 & -0.0001 & -0.0003 & 0.0001 & -0.0387 & -0.0008  & -0.0004  &  -0.0007  & -0.0391 \\
& 1000 &4$\times10^{-5}$  & -0.0001 & 0.0001 & -0.0202 & 0.0002  &  5$\times10^{-5}$ & -5$\times10^{-5}$  & -0.0187 \\
\hline
\multirow{6}{*}{Var.} &  30 & 0.3303  & 0.0867 & 0.1034 & 7$\times10^8$ &  0.3054 & 0.0874  &  0.0952 & 5$\times10^8$ \\
& 50 & 0.1465 & 0.0493 & 0.0468 & 1$\times10^8$ &  0.1377 & 0.0490  & 0.0441  & 4$\times10^7$ \\
& 100 & 0.0590 & 0.0215 & 0.0191 & 7$\times10^4$ & 0.0557  &  0.0214 & 0.0181  & 5$\times10^6$ \\
& 250 & 0.0203 & 0.0081 & 0.0069 &  0.1494 &  0.0195 &  0.0082 & 0.0065  &  0.1505 \\
& 500 & 0.0102 & 0.0039 & 0.0034 & 0.0605 & 0.0096  & 0.0039   & 0.0032  & 0.0605 \\
& 1000& 0.0050 & 0.0019 & 0.0016 & 0.0283 & 0.0047  &  0.0019 &  0.0016 & 0.0277 \\
\hline
\multirow{6}{*}{RMSE} &  30  & 0.5747 & 0.2965 & 0.3216 & 3$\times10^4$ &  0.5529 & 0.2977  & 0.3092  & 2$\times10^4$ \\
& 50 & 0.3827 & 0.2237 & 0.2165 & 1$\times10^4$ & 0.3714  & 0.2226  & 0.2103  & 6553.9 \\
& 100 & 0.2428 & 0.1470 & 0.1382 & 277.2 & 0.2360  &  0.1465  & 0.1345  & 2271.3 \\
& 250 & 0.1426 & 0.0902 & 0.0835 & 0.3965 & 0.1399  & 0.0906  &  0.0809 & 0.3979 \\
& 500 & 0.1011 & 0.0629 & 0.0583 & 0.2491 &  0.0981 &  0.0629 &  0.0566 &  0.2490\\
& 1000 & 0.0711 & 0.0445 & 0.0410 & 0.1694 & 0.0690  & 0.0443  & 0.0398  & 0.1677 \\
\hline
\end{tabular}
}

\end{center}
\end{table}

\begin{table}[!htbp]
\begin{center}
\caption{\small Simulation results: LTP $t$, $\delta=2$.}\label{table:LTPt22}
\vspace{.5em}
{\small
\begin{tabular}[h]{|c|c|c|c|c|c|c|c|c|c|}
\hline
Par. & & \multicolumn{4}{c|}{$\mu=0$, $\sigma=1$, $\gamma=0.5$, $\delta=2$} & \multicolumn{4}{c|}{$\mu=0$, $\sigma=1$, $\gamma=0.75$, $\delta=2$} \\
\hline
 & $n$ & $\hat\mu$ & $\hat\sigma$ & $\hat\gamma$ & $\hat\delta$ & $\hat\mu$ & $\hat\sigma$ & $\hat\gamma$ & $\hat\delta$ \\
\hline
\multirow{6}{*}{Bias} &  30 & -0.0434 & -0.0224 & -0.0441 & -2181.1 & -0.0263  & 0.0021  & -0.0543  & -1466.7 \\
& 50  & -0.0240 & -0.0183 & -0.0214 & -884.1 & -0.0227  & -0.0086  & -0.0308  & -409.1 \\
& 100 & -0.0087 & -0.0088 & -0.0081 & -2.3 & -0.0116  &  -0.0061 &  -0.0142 & -1.560 \\
& 250 & -0.0003 & -0.0033 & -0.0018 & -0.0896 & -0.0020  & -0.0027  & -0.0040  &  -0.0884\\
& 500 & -0.0007 & -0.0002 &  -0.0011& -0.0382 & -0.0011  & -0.0001  & -0.0021  & -0.0394 \\
& 1000 & 0.0003 & -2$\times10^{-5}$ & -0.0002 & -0.0186 & 0.0004  & 0.0001  & -0.0005  & -0.0178 \\
\hline
\multirow{6}{*}{Var.} &  30 & 0.2349 & 0.0860 & 0.0744 & 4$\times10^8$ & 0.1169  & 0.0772  & 0.0383  & 4$\times10^8$ \\
& 50 & 0.1150 & 0.0482 & 0.0373 & 2$\times10^8$ &   0.0703 & 0.0472  & 0.0226  & 2$\times10^8$ \\
& 100 & 0.0474 & 0.0214 & 0.0150 & 3$\times10^4$ &  0.0321 & 0.0211  &  0.0100 & 9640.1 \\
& 250 & 0.0164 &  0.0082 & 0.0053 & 0.1521 & 0.0105  &  0.0082 &  0.0032 & 0.1507 \\
& 500 & 0.0077 & 0.0040 & 0.0025 & 0.0610 &  0.0048 & 0.0039  & 0.0015  & 0.0617 \\
& 1000& 0.0038 & 0.0019 & 0.0012 & 0.0280 & 0.0023  &  0.0019 &  0.0007  & 0.0275 \\
\hline
\multirow{6}{*}{RMSE} &  30  & 0.4865 & 0.2941 & 0.2763 & 2$\times10^4$ & 0.3430  & 0.2779  & 0.2031  & 2$\times10^4$ \\
& 50 & 0.3400 & 0.2202 & 0.1943 & 1$\times10^4$ &  0.2661 &  0.2174 & 0.1536  & 1$\times10^4$ \\
& 100 & 0.2180 & 0.1467 & 0.1227 & 171.7 &  0.1797 & 0.1453  & 0.1007  &  98.2 \\
& 250 & 0.1283 & 0.0907 & 0.0729 & 0.4002 &  0.1024 & 0.0909  & 0.0572  & 0.3981 \\
& 500 & 0.0879 & 0.0634 & 0.0504 & 0.2500 & 0.0696  &  0.0628 & 0.0395  & 0.2515 \\
& 1000 & 0.0619 & 0.0443 & 0.0356 & 0.1685 &  0.0482 &  0.0443 &  0.0275 & 0.1670 \\
\hline
\end{tabular}
}

\end{center}
\end{table}

\begin{table}[!htbp]
\begin{center}
\caption{\small Simulation results: LTP SAS, $\delta=0.75$.}\label{table:LTPSAS1}
\vspace{.5em}
{\small
\begin{tabular}[h]{|c|c|c|c|c|c|c|c|c|c|}
\hline
Par. & & \multicolumn{4}{c|}{$\mu=0$, $\sigma=1$, $\gamma=0$, $\delta=0.75$} & \multicolumn{4}{c|}{$\mu=0$, $\sigma=1$, $\gamma=0.25$, $\delta=0.75$} \\
\hline
 & $n$ & $\hat\mu$ & $\hat\sigma$ & $\hat\gamma$ & $\hat\delta$ & $\hat\mu$ & $\hat\sigma$ & $\hat\gamma$ & $\hat\delta$ \\
\hline
\multirow{6}{*}{Bias} & 30  & 0.0140 & -148.9 & 0.0069 & -81.6 & -0.0577  & -505.4  & -0.0370  & -245.9 \\
& 50  & 0.0043 & -111.8 & 0.0007 & -57.3 & -0.0435  & -91.6  &  -0.0239 & -51.6 \\
& 100 & -0.0014 & -1.6 & -0.0003 & -0.877 & -0.0188  & -0.4154  &  -0.009 & -0.2199 \\
& 250 & 0.0001 & -0.0356 & -0.0002 & -0.0182 &  0.0032 & -0.0353  & -0.0026  & -0.0181 \\
& 500 & 0.0001 & -0.0162 & -0.0001 & -0.0085 &  -0.0015 & -0.0160  & -0.0013  & -0.0084 \\
& 1000 & 0.0005 & -0.0068 & 0.0003 & -0.0037 & 0.0004  & -0.0066  & -0.0001  & -0.0036 \\
\hline
\multirow{6}{*}{Var.} & 30 & 1.0538 & 7$\times10^6$ & 0.1913 & 2$\times10^6$ & 0.9535  & 5$\times10^8$  & 0.1758  & 1$\times10^8$ \\
& 50 & 0.4317 & 3$\times10^7$ & 0.0749 & 7$\times10^6$ &  0.3844 & 1$\times10^7$  & 0.068  & 6$\times10^6$ \\
& 100 & 0.1164 & 1$\times10^4$ & 0.0211 & 3535.2 & 0.1141  & 232.8  & 0.0207  & 69.9 \\
& 250 & 0.0366 & 0.0387 & 0.0068 & 0.0071 & 0.0346  &  0.0393  & 0.0064  & 0.0072 \\
& 500 & 0.01809 & 0.0166 & 0.0032 & 0.0030 & 0.0169  & 0.0167  & 0.0031  & 0.0030 \\
& 1000& 0.0087 & 0.0077 & 0.0015 & 0.0013 & 0.0082  & 0.0076  & 0.0015  & 0.0013 \\
\hline
\multirow{6}{*}{RMSE} & 30 & 1.0266 & 2696.3 & 0.4374 & 1449.7 & 0.9781  & 2$\times1064$  & 0.4209  & 1$\times10^4$ \\
& 50 & 0.6570 & 5815.5 & 0.2737 & 2784.7 &  0.6215 & 4139.3  & 0.2621  & 2533.5 \\
& 100 & 0.3411 & 117.2 & 0.1455 & 59.4 & 0.3383  & 15.2 & 0.1444  & 8.3 \\
& 250 & 0.1914 & 0.2001 & 0.0826 & 0.0863 & 0.1861  & 0.2015  & 0.0800  & 0.0872 \\
& 500 & 0.1345 & 0.1301 & 0.0573 & 0.0557 & 0.1300  &  0.1304 & 0.0557  & 0.0556 \\
& 1000 & 0.0934 & 0.0880 & 0.0397 & 0.0372 & 0.0910  & 0.0876  &   0.0388 & 0.0371 \\
\hline
\end{tabular}
}
\end{center}
\end{table}

\begin{table}[!htbp]
\begin{center}
\caption{\small Simulation results: LTP SAS, $\delta=0.75$.}\label{table:LTPSAS2}
\vspace{.5em}
{\small
\begin{tabular}[h]{|c|c|c|c|c|c|c|c|c|c|}
\hline
Par. & & \multicolumn{4}{c|}{$\mu=0$, $\sigma=1$, $\gamma=0.5$, $\delta=0.75$} & \multicolumn{4}{c|}{$\mu=0$, $\sigma=1$, $\gamma=0.75$, $\delta=0.75$} \\
\hline
 & $n$ & $\hat\mu$ & $\hat\sigma$ & $\hat\gamma$ & $\hat\delta$ & $\hat\mu$ & $\hat\sigma$ & $\hat\gamma$ & $\hat\delta$ \\
\hline
\multirow{6}{*}{Bias} &  30 & -0.0817 & -593.2 & -0.0647 & -330.2 & -0.0450  & -49.5  & -0.0692  & -27.9 \\
& 50  & -0.0712 & -71.1 & -0.0429 & -37.9 & -0.0584  & -11.3  & -0.0502  & -5.8 \\
& 100 & -0.0338 & -0.5098 & -0.0183 & -0.2595 & -0.0372  & -0.2321  & -0.0252  & -0.1262 \\
& 250 & -0.0068 & -0.0351 & -0.0050 & -0.0182 & -0.0090  & -0.0344  & -0.0072  & -0.0180 \\
& 500 & -0.0033 & -0.0161 & -0.0027 & -0.0085 & -0.0038  &  -0.0171 & -0.0034  & -0.0088 \\
& 1000 & -0.0001 & -0.0065 & -0.0005 & -0.0036 & 0.0007  & -0.0068  & -0.0012  & -0.0036 \\
\hline
\multirow{6}{*}{Var.} & 30 & 0.6246 & $4\times10^8$ & 0.1188 & 1$\times10^8$ & 0.0264  & 3$\times 10^6$  &  0.0053 & 1$\times 10^6$ \\
& 50 & 0.2669 & $9\times10^6$ & 0.0497 & 2$\times10^6$ &  0.1293 &  2.6$\times10^5$ & 0.0256  & 5$\times10^4$ \\
& 100 & 0.0982 & 856.7 & 0.0179 & 215.0 &  0.0632 &  96.4 &  0.0115 & 29.9 \\
& 250 & 0.0295 & 0.0390 & 0.0053 & 0.0072 & 0.0195  &  0.0391 & 0.0034 & 0.0072 \\
& 500 & 0.0141 & 0.0166 & 0.0025 & 0.0030 & 0.0086  & 0.0167  & 0.0015  &  0.0029\\
& 1000& 0.0067 & 0.0076 & 0.0012 & 0.0013 &  0.0040 &  0.0077 & 0.0007  & 0.0013 \\
\hline
\multirow{6}{*}{RMSE} &  30  & 0.9453 & 2$\times10^4$ & 0.3507 & 1$\times10^4$ & 0.5163  &  1739.4 &  0.2408 & 1016.1 \\
& 50 &0.5215  & 3085.5 & 0.2270 & 1618.6 & 0.3642  & 513.8  & 0.1678  & 241.8 \\
& 100 & 0.3153 & 29.2 &  0.1352 & 14.6 & 0.2542  & 9.8  & 0.1105  & 5.4 \\
& 250 & 0.1719 & 0.2007 & 0.0735 & 0.0869 & 0.1401  & 0.2009  & 0.0594  & 0.0871 \\
& 500 & 0.1188 & 0.1300 & 0.0507 & 0.0554 & 0.0933  & 0.1303  &  0.0396 & 0.0554 \\
& 1000 & 0.0823 & 0.0879 & 0.0350 & 0.0373 & 0.0640  &  0.0881 & 0.0272  & 0.0375 \\
\hline
\end{tabular}
}
\end{center}
\end{table}

\begin{table}[!htbp]
\begin{center}
\caption{\small Simulation results: LTP Normal.}\label{table:LTPNR}
\vspace{.5em}
{\small
\begin{tabular}[h]{|c|c|c|c|c|c|c|}
\hline
\multicolumn{7}{|c|}{$\beta_1=1$, $\beta_2=2$, $\beta_3=3$, $\sigma=0.25$, $\gamma=0$} \\
\hline
 & $n$ & $\hat\beta_1$ & $\hat\beta_2$& $\hat\beta_3$ &  $\hat\sigma$ & $\hat\gamma$   \\
\hline
\multirow{3}{*}{Bias} &  100 & 0.0060 & -0.0118 &  -0.0113 & 0.0061 & 0.0025 \\
& 250 & 0.0027 &  -0.0041 & -0.0051 & 0.0023 &  0.0013 \\
& 500 & 0.0011 & -0.0008 & -0.0018 & 0.0012 & 0.0011\\
\hline
\multirow{3}{*}{Var.} &  100 & 0.0128 & 0.0290 & 0.0387 & 0.0004 & 0.0502 \\
& 250 & 0.0039 & 0.0107 & 0.0141 & 0.0002 & 0.0135 \\
& 500 & 0.0019 & 0.0052 & 0.0069 & 0.0001 &  0.0062\\
\hline
\multirow{3}{*}{RMSE} &  100 & 0.1134 & 0.1708 & 0.1971 & 0.0212 & 0.2242 \\
& 250 & 0.0628 & 0.1037 & 0.1190 & 0.0132 & 0.1165 \\
& 500 & 0.0436 & 0.0724 & 0.0836 & 0.0093 & 0.0792\\
\hline
 \multicolumn{7}{|c|}{$\beta_1=1$, $\beta_2=2$, $\beta_3=3$, $\sigma=0.25$, $\gamma=0.25$} \\
\hline
\multirow{3}{*}{Bias} &  100 & -0.0036 & -0.0120 & -0.0105 & 0.0067 & -0.0293 \\
& 250 & 1$\times 10^{-5}$ & -0.0004 & -0.0004 & 0.0024 & -0.0008 \\
& 500 & 4$\times 10^-5$ & -0.0011 & -0.0019 & 0.0012 & -0.0029\\
\hline
\multirow{3}{*}{Var.} &  100 & 0.0127 & 0.0284 & 0.0364 & 0.0004 & 0.0525\\
& 250 & 0.0037 & 0.0101 & 0.0133 & 0.0002 & 0.0128\\
& 500 & 0.0017 & 0.0049 & 0.0064 & 8$\times10^{-5}$ & .0058 \\
\hline
\multirow{3}{*}{RMSE} &  100 & 0.1128 & 0.1690 & 0.1913 & 0.0209 & 0.2311 \\
& 250  & 0.0611 & 0.1006 & 0.1157 & 0.0129 & 0.1136 \\
& 500 & 0.0421 & 0.0705 & 0.0805 & 0.0089 & 0.0766 \\
\hline
 \multicolumn{7}{|c|}{$\beta_1=1$, $\beta_2=2$, $\beta_3=3$, $\sigma=0.25$, $\gamma=0.5$} \\
\hline
\multirow{3}{*}{Bias} &  100 & -0.0150 & -0.0073 & -0.0087 & 0.0077 & -0.0648 \\
& 250 & -0.0032 & -0.0027 & -0.0040 & 0.0025 & -0.0170 \\
& 500  & -0.0015 & -0.0003 & -0.0011 & 0.0013 & -0.0075 \\
\hline
\multirow{3}{*}{Var.} &  100 & 0.0112 & 0.0252 & 0.0329 & 0.0004 & 0.04814 \\
& 250 & 0.0033 &  0.0084 & 0.0110 & 0.0002 &  0.0117\\
& 500 & 0.0015 &   0.0040 & 0.0054 &  $8\times10^-5$& 0.0049 \\
\hline
\multirow{3}{*}{RMSE} &  100 & 0.1071 & 0.1592 & 0.1816 & 0.0210 &  0.2288\\
& 250 & 0.0580 & 0.0920 & 0.1053 & 0.0125 & 0.1098 \\
& 500 & 0.0386 & 0.0633 & 0.0735 & 0.0087 & 0.0708 \\
\hline
\end{tabular}
}
\end{center}
\end{table}

\begin{table}[!htbp]
\begin{center}
\caption{\small Simulation results: LTP SAS.}\label{table:LTPSASR}
\vspace{.5em}
{\small
\begin{tabular}[h]{|c|c|c|c|c|c|c|c|}
\hline
\multicolumn{8}{|c|}{$\beta_1=1$, $\beta_2=2$, $\beta_3=3$, $\sigma=0.25$, $\gamma=0$, $\delta = 0.75$} \\
\hline
 & $n$ & $\hat\beta_1$ & $\hat\beta_2$& $\hat\beta_3$ &  $\hat\sigma$ & $\hat\gamma$ & $\hat\delta$   \\
\hline
\multirow{3}{*}{Bias} & 100 &  0.0025  & -0.0196 & -0.0240 & -0.0580 & -0.0118 &  -0.12704  \\
& 250 &  0.0017 & -0.0063 & -0.0086 & -0.0091 & -0.0028 &   -0.0210  \\
& 500 & 0.0004  & -0.0023 & -0.0033 & -0.0038 & -0.0015 &  -0.0094  \\
\hline
\multirow{3}{*}{Var.} & 100 & 0.0250  & 0.0658 & 0.0865 & 0.2691 &0.0466  &  1.23  \\
& 250 & 0.0063  & 0.0233 & 0.0301 & 0.0038 & 0.0102 &   0.0122  \\
& 500 & 0.0030  & 0.0112 & 0.0145 & 0.0014 & 0.0046 &  0.0044   \\
\hline
\multirow{3}{*}{RMSE} & 100 & 0.1582  & 0.2573 & 0.2951 & 0.5220 & 0.2163 &  1.11   \\
& 250 &  0.0798 & 0.1529 & 0.1737 & 0.0624 & 0.1011 &  0.1127   \\
& 500 & 0.0547  & 0.1061 & 0.1207 & 0.0383 &0.0682  &  0.0674   \\
\hline
\multicolumn{8}{|c|}{$\beta_1=1$, $\beta_2=2$, $\beta_3=3$, $\sigma=0.25$, $\gamma=0.25$, $\delta = 0.75$} \\
\hline
\multirow{3}{*}{Bias} & 100 & -0.0073  & -0.0221 & -0.0203 & -0.0470 & -0.0342 &  -0.1041   \\
& 250 &  0.0005 & -0.0075 & -0.0090 & -0.0082 & -0.0078 &  -0.0193   \\
& 500 & -0.0002  & -0.0024 & -0.0032 & -0.0036 & -0.0039 &  -0.0089   \\
\hline
\multirow{3}{*}{Var.} & 100 & 0.0226  & 0.0643 & 0.0806 & 0.0977 & 0.0426 &  0.3989   \\
& 250 &  0.0060 & 0.0219 & 0.0283 & 0.0033 &  0.0094 &   0.0103  \\
& 500 & 0.0028  & 0.0106 & 0.0135 & 0.0013 & 0.0044 &  0.0039   \\
\hline
\multirow{3}{*}{RMSE} & 100 & 0.1506  & 0.2546 & 0.2847 & 0.3162 & 0.2094  & 0.6401    \\
& 250 & 0.0775  & 0.1482 & 0.1685 & 0.0585 & 0.0972 &  0.1036   \\
& 500 & 0.0533  & 0.1033 & 0.1165 & 0.0368 & 0.0664 &  0.0637   \\
\hline
 \multicolumn{8}{|c|}{$\beta_1=1$, $\beta_2=2$, $\beta_3=3$, $\sigma=0.25$, $\gamma=0.5$, $\delta = 0.75$} \\
\hline
\multirow{3}{*}{Bias}& 100 &  -0.0122 & -0.0156 & -0.0191 & -0.0336 & -0.0486 &  -0.0782   \\
& 250 &  -0.0013 & -0.0058 & -0.0092 & -0.0075 & -0.0127 &  -0.0181   \\
& 500 &  -0.0011 & -0.0017 & -0.0030 & -0.0034 & -0.0064 &  -0.0085   \\
\hline
\multirow{3}{*}{Var.} & 100 & 0.0174  & 0.0559 & 0.0700 & 0.0419 & 0.0325 &  0.1593   \\
& 250 & 0.0051  & 0.0182 & 0.0237 & 0.0028 &  0.0078 &  0.0087  \\
& 500 & 0.0023  & 0.0087 & 0.0114 & 0.0012 & 0.0034 &  0.0035   \\
\hline
\multirow{3}{*}{RMSE} & 100 & 0.1328  & 0.2369 & 0.2654 & 0.2075 & 0.1868 &   0.4067  \\
& 250 & 0.0720  & 0.1353 & 0.1543 & 0.0543 & 0.0894 &  0.0952   \\
& 500 & 0.0489  & 0.0935 & 0.1069 & 0.0350 & 0.0594 &  0.0598   \\
\hline
\end{tabular}
}
\end{center}
\end{table}

\clearpage
\section{Applications}\label{NumericalExamples}

In this section, we present several medical applications with real data that illustrate the performance and usefulness of the proposed distributions. Throughout, we adopt the epsilon-skew parameterisation $\{a(\gamma),b(\gamma)\}=\{1-\gamma,1+\gamma\}$, $\gamma \in (-1,1)$, for the LTP distributions.

\subsection{Example 1: Nerve data}

\begin{figure}[h!]
\begin{center}
\begin{tabular}{c c}
\psfig{figure=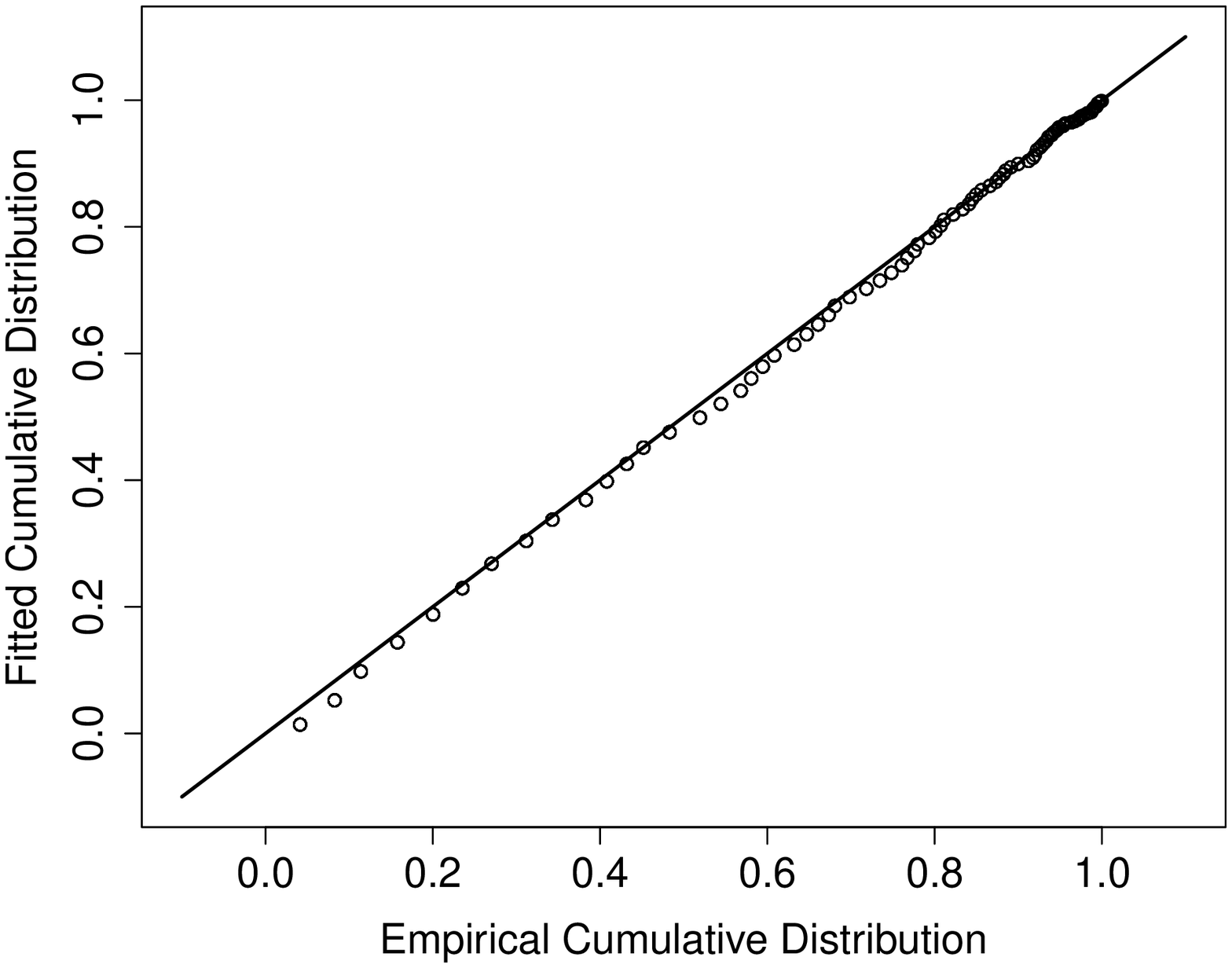,  height=6cm, width = 6cm}  &
\psfig{figure=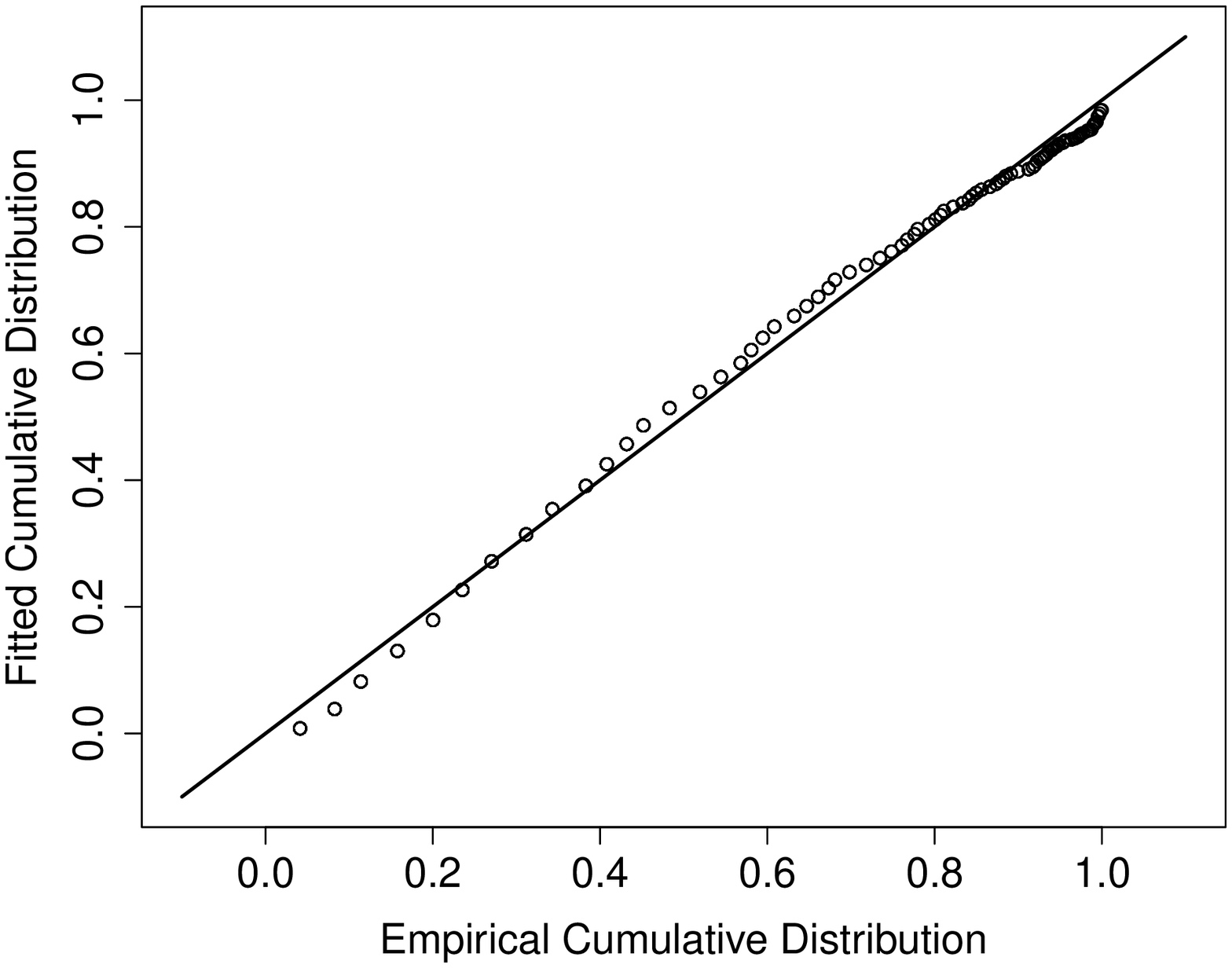,  height=6cm, width = 6cm}\\
(a) & (b)\\
\psfig{figure=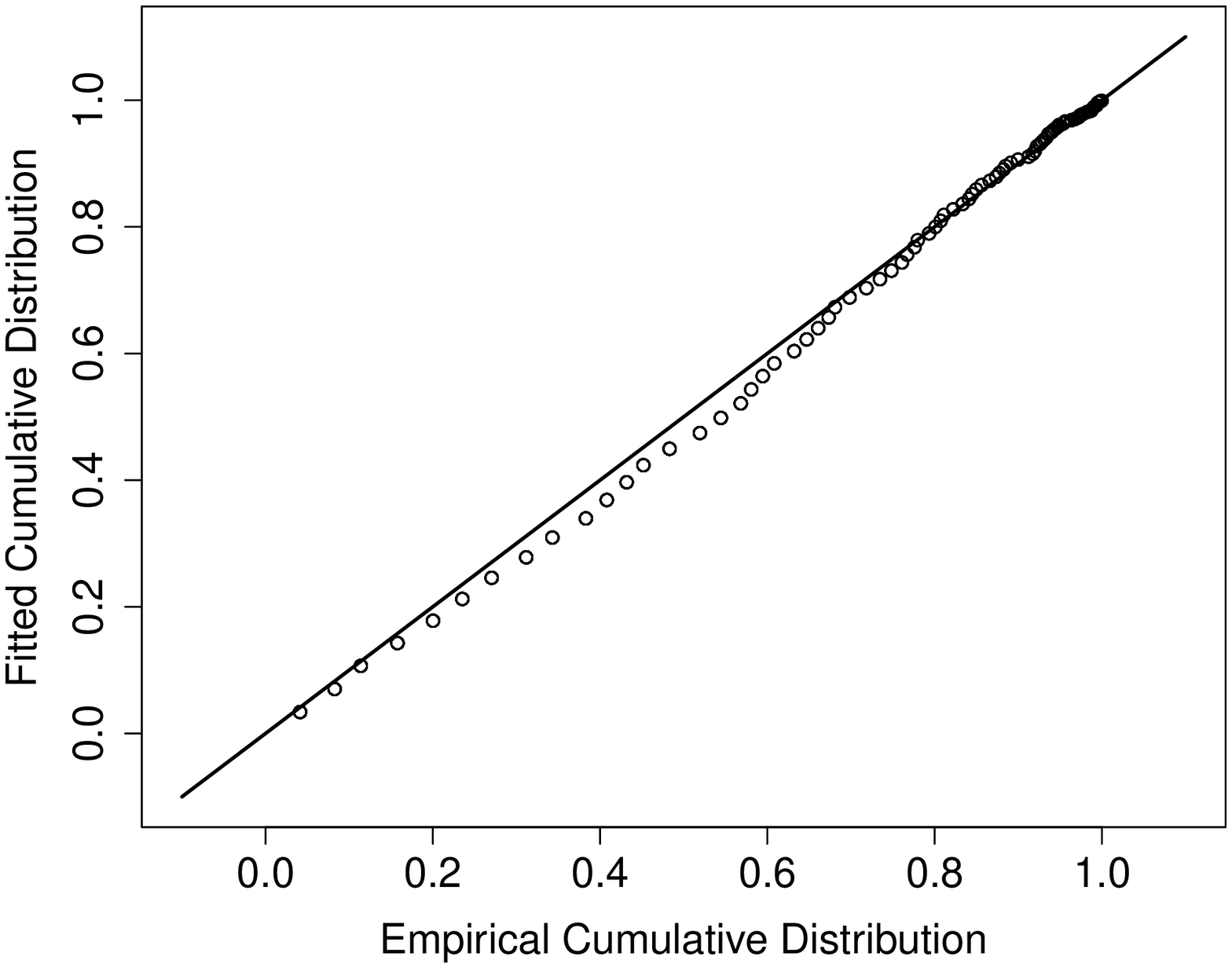,  height=6cm, width = 6cm}  &
\psfig{figure=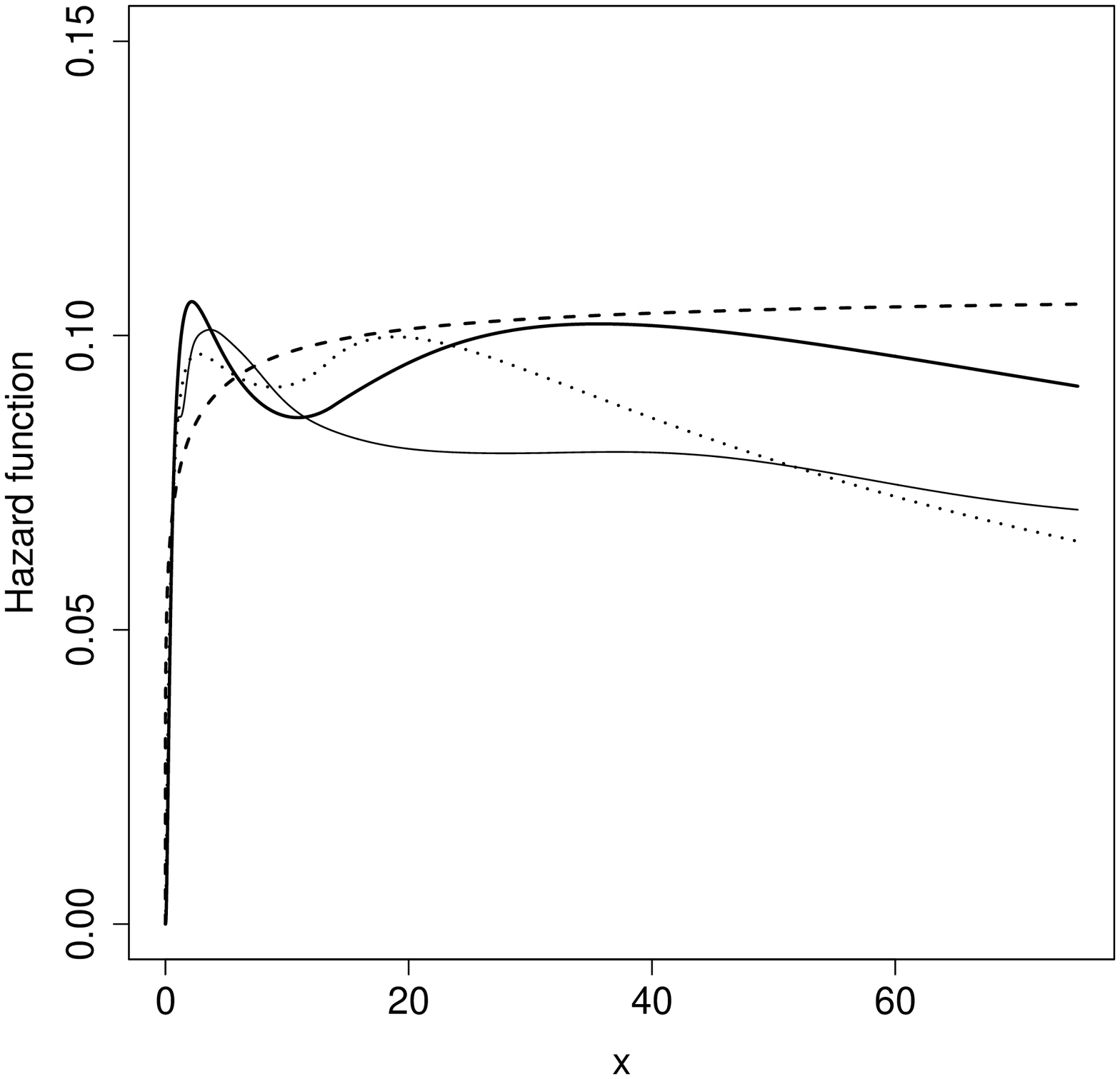,  height=6cm, width = 6cm}\\
(c) & (d)
\end{tabular}
\end{center}
\caption{\small Nerve data: (a) LTP SAS probability plot; (b) lognormal probability plot; (c) Weibull probability plot; and (d) Fitted hazard functions. LTP SAS (bold line), LTP Normal (dotted line), Gamma (dashed line), and kernel estimator (continuous line).}
\label{fig:NFM}
\end{figure}

In our first example we analyse the data set reported in \cite{CL66} which contains $n=799$ observations rounded to the nearest half in units of 1/50 second, which correspond to the time between $800$ successive pulses along a nerve fibre. We consider two baseline distributions $s$ in (\ref{LTPPDF}): a Student $t$ density with $\delta>0$ degrees of freedom (LTP $t$), and a symmetric sinh-arcsinh density \citep{JP09,R15} (LTP SAS). The choice for these two baseline densities is motivated as follows. The Student-$t$ distribution is a parametric family of distributions with heavier tails than the normal ones; having the normal distribution as a limit case when $\delta \rightarrow\infty$. The behaviour of the tails of the Student-$t$ density is polynomial. On the other hand, the symmetric sinh-arcsinh density (reported in Appendix A) is a parametric density function which contains a parameter that controls the tail behaviour. This distribution can capture tails heavier or lighter than those of the normal density ($\delta\lessgtr 1$), being the normal distribution a particular case ($\delta=1$). The tails of the symmetric sinh-arcsinh density are lighter than any polynomial \citep{JP09}. Therefore, with these two choices of the baseline density we can cover a wide range of tail behaviours. Moreover, with the additional shape parameter $\gamma$ we also cover a wide range of shapes around the shoulders of the density. Table \ref{table:NerveMLE} shows the MLEs and the Akaike information criterion (AIC) associated to these models as well as some natural competitors. We also report the estimators of the LTP Normal and the lognormal distributions, which are particular cases of the LTP SAS model. The AIC favours the LTP SAS model overall, closely followed by the LTP Normal. The MLE of the parameter $\delta$ in the LTP SAS model is larger than one, indicating that the data favour a model with lighter tails than those of the lognormal distribution. The 95\% confidence intervals for the parameters $\gamma$ and $\delta$ (obtained as the 0.147-level profile likelihood intervals, see \cite{K85}) in the LTP SAS model are $(0.31,0.53)$ and $(1.02, 1.56)$, respectively. It is worth noticing that the confidence interval for $\delta$ only include values greater than one, which are associated to tails lighter than normal. Figures \ref{fig:NFM}a--\ref{fig:NFM}c show the probability plots and hazard functions corresponding to the LTP SAS, lognormal, and Weibull models, which visually illustrates the fit of these models. From Figure \ref{fig:NFM}d We can observe that the fitted Gamma model produces an increasing hazard function, while the LTP SAS model produces a non monotonic hazard function with decreasing tail. This behaviour coincides with that of the fitted kernel estimation of the hazard function (which was obtained using lognormal kernels).

\begin{table}[h!]
\begin{center}
\caption{\small Nerve data: Maximum likelihood estimates, AIC (best value in bold).}
\label{table:NerveMLE}
\vspace{.5em}
\begin{tabular}[h]{|c|c|c|c|c|c|}
\hline
Model & $\hat\mu$ & $\hat\sigma$ & $\hat\gamma$ & $\hat\delta$ & AIC\\
\hline
LTP $t$ &  2.59 & 1.04  & 0.40 & 111.90 & 5401.80 \\
LTP SAS & 2.63  &  1.39 & 0.42 & 1.22 & {\bf 5395.71} \\
LTP Normal & 2.59  & 1.05  & 0.40 & -- &  5398.45 \\
Log-normal &  1.91  & 1.08  & -- & -- & 5443.70 \\
Weibull & --  & (scale) 11.27 & (shape) 1.08 & -- & 5415.40  \\
Gamma &  -- & (scale) 9.31   & (shape) 1.17  & --  & 5411.11 \\
\hline
\end{tabular}

\end{center}
\end{table}
%

\pagebreak

\subsection{Example 2: PBC data}

In this section, we analyse the popular Mayo primary biliary cirrhosis (PBC) data, reported in Appendix D from \cite{FH91}, in order to illustrate the performance of the proposed distributions in the context of AFT models. This data set contains information about the survival time and prognostic factors for 418 patients in a study conducted at Mayo Clinic between 1974 and 1984. The survival times are reported in days together with an indicator variable associated to the status of the patient at the end of the study (0/1/2 for censored, transplant, dead). \cite{J03} fitted, using a semiparametric method, an AFT model with five covariates: age (in years), logarithm of the serum albumin (in mg/dl), logarithm of the serum bilirubin (in mg/dl), edema, and logarithm of the prothrombin time (in seconds). Similarly, \cite{D10} reports the semiparametric estimators of the AFT model with an intercept parameter as follows: $(8.692,-0.025,1.498,-0.554,-0.904,-2.822)$. We consider a maximum likelihood estimation approach of the AFT model (\ref{AFTModel}) containing an intercept and LTP $t$ errors with parameters $(0, \sigma_{\varepsilon}, \gamma_{\varepsilon}, \delta_{\varepsilon})$. The estimators and the AIC values are reported in Table \ref{table:PBCMLE}. We can see that the estimators obtained for the model with LTP $t$ and Log-$t$ models are close to those reported by \cite{D10} using a semiparametric method. The AIC values favour the models with LTP-$t$ and Log Student-$t$ (Log-$t$) errors. However, these values do not provide strong evidence to distinguish between the two models, and therefore the model choice deserves further investigation. The MLE of the skewness parameter $\gamma_{\varepsilon}$ is relatively far from zero in the Log-$t$ model. However, the inclusion of this parameter produces little effect in the estimation of the degrees of freedom $\delta_{\varepsilon}$ and the regression parameters. The 95\% confidence interval for $\gamma_{\varepsilon}$ in the LTP-$t$ model is $(-0.374,0.167)$, which does not rule out the value $\gamma_{\varepsilon}=0$ as a likely value of the parameter. Then, a parsimony argument favours the model with log-Student $t$ errors (Log-$t$) in this case. Moreover, we can observe that the MLEs of $\gamma_{\varepsilon}$ in the LTP $t$ and LTP Normal model have different signs. The 95\% confidence interval for $\gamma_{\varepsilon}$ in the LTP Normal model is $(-0.072,0.493)$ (which indicates that $\gamma_{\varepsilon}=0$ is an unlikely value of the parameter). The reason for this difference is that the data seem to favour a model with heavier tails than normal. The lack of flexibility in the tails and the presence of extreme observations affect the estimation of the shape parameter $\gamma_{\varepsilon}$ in the LTP Normal model by pulling out this estimator in the opposite direction. This emphasises the importance of assessing the type of flexibility required for properly modelling the data.


\begin{table}[h!]
\begin{center}
\caption{\small PBC data: Maximum likelihood estimates, AIC (best values in bold).}
\label{table:PBCMLE}
\vspace{.5em}
\begin{tabular}[h]{|c|c|c|c|c|}
\hline
Model & LTP $t$ & LTP Normal &  Log-$t$& Log-normal \\\hline
Intercept & 7.704 & 7.518 & 7.539 & 7.731 \\
Age  & -0.026 & -0.026 & -0.027 & -0.025 \\
log(Albumin)  & 1.552 & 1.529 & 1.554 & 1.472\\
log(Bilirubin)  & -0.587 & -0.620  & -0.595 & -0.606 \\
Edema  & -0.762 & -0.710 & -0.706 & -0.840 \\
log(Protime)  & -2.464 & -2.189 & -2.313 & -2.371 \\
$\sigma_{\varepsilon}$  &  0.773 & 0.908 & 0.770 & 0.973 \\
$\gamma_{\varepsilon}$  & -0.133 & 0.190 & -- & --\\
$\delta_{\varepsilon}$ & 4.446 & -- & 5.602 & -- \\\hline
AIC & {\bf 635.019} & 642.318 & {\bf 633.702} & 642.222 \\
\hline
\end{tabular}

\end{center}
\end{table}

\pagebreak
\subsection{Example 3: NCCTG Lung Cancer Data}\label{Lung}

In this section, we revisit the popular NCCTG Lung Cancer Data. This data set contains the survival times of $n=227$ patients (the total number of patients is 228 but we have removed one patient with a missing covariate, for the sake of simplicity) with advanced lung cancer from the North Central Cancer Treatment Group. The goal of this study was to compare the descriptive information from a questionnaire applied to a group of patients against the information obtained by the patient's physician, in terms of prognostic power \citep{L94}. We fit an AFT model with three covariates ``age'' (in years),``sex'' (Male=1 Female=2), ``ph.ecog'' (ECOG performance score, 0=good--5=dead) as well as an intercept, with different choices for the distribution of the errors in (\ref{AFTModel}). Table \ref{table:LungMLE} shows the MLEs associated to each of these models together with the AIC values. The AIC favours the model with LTP logistic errors, closely followed by the model with LTP SAS errors. One explanation for this is that the estimators of the LTP SAS model indicate tails heavier than normal ($\hat\delta=0.6674$), which is a tail behaviour naturally captured by the LTP logistic distribution without additional shape parameters that control the tail. The 95\% confidence intervals for $\gamma$ and $\delta$ in the LTP SAS model are $(-0.05,0.60)$ and $(0.46,0.86)$, respectively, while the corresponding confidence interval for the parameter $\gamma$ in the LTP logistic model is $(0.16, 0.62)$.

\begin{table}[h!]
\begin{center}
\caption{\small NCCTG Lung Cancer data: Maximum likelihood estimates, AIC (best value in bold).}
\label{table:LungMLE}
\vspace{.5em}
\begin{tabular}[h]{|c|c|c|c|c|c|}
\hline
Model & LTP SAS & LTP Normal & Log-normal&  LTP logistic & Log-logistic \\\hline
Intercept & 6.2077 & 6.9505 & 6.4940 & 6.5538  & 5.9500 \\
Age   & -0.0068 & -0.0149 & -0.0191 & -0.0100 & -0.0082 \\
Sex   & 0.4614 & 0.4259 & 0.5219 & 0.4243 & 0.4857 \\
ph.ecog   & -0.3824 & -0.3121 & -0.3557 & -0.3541 & -0.4042 \\
$\sigma$   & 0.4639 & 0.8835 & 1.0286 & 0.4847 & 0.5360 \\
$\gamma$   & 0.3095 & 0.5051 & -- &  0.4083 & -- \\
$\delta$   & 0.6674 & -- & -- & -- & -- \\\hline
AIC  & 538.2100 & 545.9405 & 563.8323  & {\bf 536.0556} & 545.0486 \\
\hline
\end{tabular}

\end{center}
\end{table}

\begin{figure}[h!]
\begin{center}
\begin{tabular}{c}
\psfig{figure=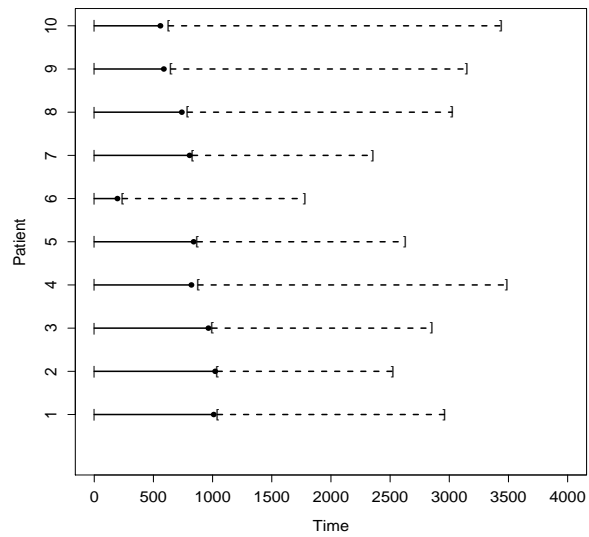,  height=8cm,width=8cm}
\end{tabular}
\end{center}
\caption{\small NCCTG Lung Cancer data: Prediction intervals for the remaining life in days. The solid line indicates the survival time in days, while the dashed line corresponds to a 90\% prediction interval for the remaining life.}
\label{fig:PredInt}
\end{figure}

It is sometimes of interest to obtain information about the remaining life of individual cancer patients. This information is used for future planning of health care, which is of financial and medical importance. Specifically, the probability that patient $i$ survives until time $t$, given that he/she was alive at time $t_i$ is given by,
\begin{eqnarray}\label{RemainingLife}
G(t\vert t_i;\bm{\theta}) = {\mathbb P}(T\leq t\vert T> t_i) = \dfrac{G(t;\bm{\theta})-G(t_i;\bm{\theta})}{1-G(t_i;\bm{\theta})},\,\,\, t\geq t_i,
\end{eqnarray}
where $G$ is the distribution under the model of interest. For an AFT model, the parameter $\bm{\theta}$ contains both the regression parameters as well as the parameters of the distribution of the errors. The simplest way to obtain an estimator of this probability consists of plugging in the MLE of $\bm{\theta}$ in (\ref{RemainingLife}). The 100(1-$\alpha$)\% prediction interval \citep{H09} for a patient that survived until time $t_i$ is $[T_i^L,T_i^R]$, which satisfies $G(T_i^L\vert t_i;\bm{\theta})=\alpha_1$ and $G(T_i^R\vert t_i;\bm{\theta})=1-\alpha_2$, with $\alpha_1 + \alpha_2=\alpha$. In our application we choose $\alpha_1=\alpha_2=0.05$, and we centre the prediction intervals at the mean of the regression model with LTP logistic errors. Figure \ref{fig:PredInt} shows the 90\% prediction interval for the remaining life for 10 censored patients.

\begin{figure}[h!]
\begin{center}
\begin{tabular}{c}
\psfig{figure=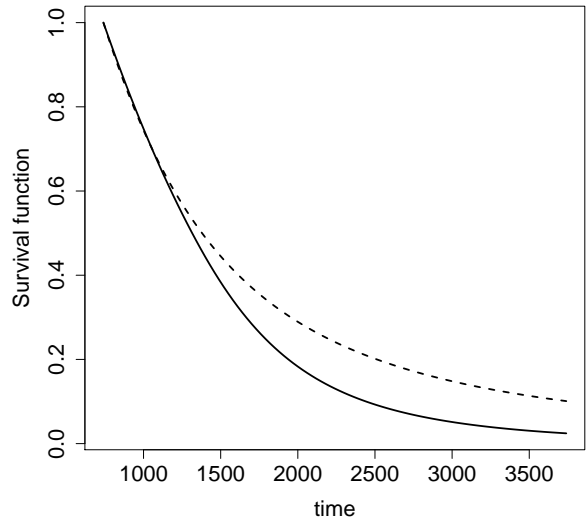,  height=8cm,width=8cm}
\end{tabular}
\end{center}
\caption{\small NCCTG Lung Cancer data: survival function of the remaining life in days. The continuous line represents the survival function of the remaining life for the LTP logistic model, and the dashed line represents the corresponding survival function for the logistic model.}
\label{fig:Patient3}
\end{figure}

We can observe from Table \ref{table:LungMLE} that the estimators of the regression parameters are very similar for the different choices of the distribution of the errors. At first glance, one might think that the choice of the distribution of the residual errors has little impact on the inference. However, if the interest in on predicting the remaining life of censored patients, we may obtain different intervals for different models. For instance, Figure \ref{fig:Patient3} shows how different the survival functions of the remaining life for a particular censored patient, associated to the models with LTP logistic and logistic errors, can be. This emphasises the importance of the correct specification of the distribution of the residual errors.

\pagebreak
\section{Discussion}

We have proposed a flexible class of parametric distributions (LTP) with positive support that can be used for the modelling of survival data. We have shown that some members of this class of distributions represent a flexible extension of the classical choices such as the lognormal, log-logistic, and log Student-$t$ distributions. The genesis of LTP distributions allows the user to play with different baseline log-symmetric distributions in order to properly model the tail behaviour of the data. These distributions can be used to produce models that are robust to departures from the assumption of log-symmetry. Moreover, LTP distributions preserve the ease of use of the baseline log-symmetric distribution. For instance, in models that assume lognormality, a LTP-normal can be implemented with virtually the same parsimony level. In practice, we recommend to conduct a model selection between 4-parameter LTP models and the corresponding 3- and 2-parameter submodels. Given that the parameters of LTP distributions are easily-interpreted, this model selection provides information about the features favoured by the data, such as asymmetry and tail behaviour, providing in turn more insights on the phenomenon of interest. Model selection between these nested models can be conducted either using AIC or the likelihood ratio test. The good behaviour of the MLE in this family can be established by appealing to the literature on the study of inferential properties of the family of two-piece distributions, which are linked to the proposed models via a logarithmic transformation. Confidence intervals for the model parameters can be obtained by using the profile likelihood. This approach avoids relying on asymptotic results, such as normal confidence intervals (standard errors), that may not be accurate for small or moderate sample sizes.

We conclude by pointing out possible extensions of our work. Multivariate extensions of the family of LTP distributions can be produced by using copulas. This approach has the advantage of separating the role of the parameters that control the shape of the distribution of the marginals and the dependencies between the marginals. As discussed in \cite{H09}, the plug-in estimators considered in Section \ref{Lung} may produce prediction intervals of the remaining life with a smaller coverage probability. The calibration of these intervals to improve their coverage in the context of LTP models represents an interesting research line.

\section*{Acknowledgements}
We thank the Editor and two referees for their very helpful comments. FJR gratefully acknowledges research support from EPSRC grant EP/K007521/1.


\section*{Appendix A: Some density functions}
Throughout we use the notation $t=\dfrac{x-\mu}{\sigma}$.
\begin{itemize}
\item The symmetric sinh-arcsinh distribution \citep{JP09}:
\begin{eqnarray*}
f(x;\mu,\sigma,\delta) = \dfrac{\delta}{\sigma}\phi\left[\sinh\left(\delta \operatorname{arcsinh}\left(t\right)\right)\right] \dfrac{\cosh\left(\delta \operatorname{arcsinh}\left(t\right)\right)}{\sqrt{1+t^2}},
\end{eqnarray*}
where $\delta>0$ controls the tails of the density, and $\phi$ is the standard normal density function. Note that for $\delta=1$, this density corresponds to the normal density.

\item The exponential power distribution:
\begin{eqnarray*}
f(x;\mu,\sigma,\delta) = \dfrac{\delta}{2\sigma\Gamma(1/\delta)}\exp\left(-\vert t \vert^\delta\right).
\end{eqnarray*}

\noindent where $\Gamma(\cdot)$ is the gamma function. This family contains the Laplace distribution for $\delta=1$ and the normal distribution (with variance $\sigma^2/2$) for $\delta=2$.

\item The Student-$t$ distribution:
\begin{eqnarray*}
f(x;\mu,\sigma,\delta) = \dfrac{\Gamma\left(\dfrac{\delta+1}{2}\right)}{\sigma\sqrt{\pi\delta } \Gamma\left(\dfrac{\delta }{2}\right)} \left(1+\dfrac{t^2}{\delta}\right)^{-\dfrac{\delta
   +1}{2}},
\end{eqnarray*}
\noindent where $\Gamma(\cdot)$ is the gamma function.

\item The logistic distribution:
\begin{eqnarray*}
f(x;\mu,\sigma) = \dfrac{1}{\sigma} \dfrac{\exp(-t)}{[1+\exp(-t)]^2}.
\end{eqnarray*}

\end{itemize}


\begin{thebibliography}{9}
\bibitem[Azzalini and Genton(2008)]{AG08} Azzalini, A. and Genton, M. G. (2008). Robust likelihood methods based on the skew-$t$ and related distributions. {\sl International Statistical Review} 76: 106--129.
\bibitem[Arellano-Valle et al.(2005)]{A05} Arellano-Valle, R. B., G{\'o}mez, H. W. and Quintana, F. A.  (2005). Statistical inference for a general class of asymmetric distributions. {\sl Journal of Statistical Planning and Inference} 128: 427--443.
\bibitem[Barros et al.(2008)]{B08} Barros, M., Paula, G. A. and Leiva, V. (2008). A new class of survival regression models with heavy-tailed errors: robustness and diagnostics. {\sl Lifetime Data Analysis} 14:316-–332.
\bibitem[Cox and Lewis(1966)]{CL66} Cox D. R. and Lewis P. A. W. (1966). {\sl The statistical analysis of series of events}. Methuem, London.
\bibitem[Ding(2010)]{D10} Ding, Y. (2010). Some new insights about the accelerated failure time model. PhD thesis, The University of Michigan.
\bibitem[Fern{\'a}ndez and Steel(1998)]{FS98} Fern{\'a}ndez, C. and Steel, M. F. J. (1998).  On Bayesian modeling of fat tails and skewness. {\sl Journal of the American Statistical Association} 93: 359--371.
\bibitem[Ferreira and Steel(2006)]{FS06} Ferreira, J. T. A. S. and Steel, M. F. J. (2006). A constructive representation of univariate skewed distributions. {\sl Journal of the American Statistical Association} 101: 823--829.
\bibitem[Fleming and Harrington(1991)]{FH91} Fleming, T. R. and Harrington, D. P. (1991). {\sl Counting Processes and Survival Analysis}. Wiley: New York.
\bibitem[Hong et al.(2009)]{H09} Hong, Y., Meeker, W. Q. and McCalley, J. D. (2009). Prediction of the remaining life of power transformers based on left truncated and right censored lifetime data. {\sl Annals of Applied Statistics} 2: 857-879.
\bibitem[Jin et al.(2003)]{J03} Jin, Z., Lin, D. Y., Wei, L. J., and Ying, Z. (2003). Rank--based inference for the accelerated failure time model. {\sl Biometrika} 90: 341--353.
\bibitem[Jones and Anaya-Izquierdo(2010)]{JA10} Jones, M. C., and Anaya-Izquierdo K. (2010). On Parameter Orthogonality in Symmetric and Skew Models. {\sl Journal of Statistical Planning and Inference} 141: 758--770.
\bibitem[Jones and Pewsey(2009)]{JP09} Jones, M. C., and Pewsey A. (2009). Sinh-arcsinh Distributions. {\sl Biometrika} 96: 761--780.
\bibitem[Kalbfleisch(1985)]{K85} Kalbfleisch, J. G. (1985). {\sl Probability and Statistical Inference: Volume 2: Statistical Inference} (Second Edition). Springer-Verlag, New York.
\bibitem[Kom{\'a}rek and Lesaffre(2008)]{KL08} Kom{\'a}rek, A. and Lesaffre, E. (2008). Bayesian accelerated failure time model with multivariate doubly interval--censored data and flexible distributional assumptions. {\sl Journal of the American Statistical Association} 103: 523--533.
\bibitem[Kotz et al.(2001)]{K01} Kotz, S., Kozubowski, T. J. and Podg{\'o}rski, K. (2001). {\sl The Laplace Distribution and Generalizations: A Revisit with Applications to Communications, Economics, Engineering, and Finance}. Birkhauser, Boston.
\bibitem[Lawless(2003)]{L03} Lawless, J. F. (2003). {\sl Statistical Models and Methods for Lifetime Data} (2nd. Edition). John Wiley \& Sons. Inc., New Jersey.
\bibitem[Loprinzi et al.(1994)]{L94}  Loprinzi C. L., Laurie, J. A., Wieand, H. S., Krook, J. E., Novotny, P. J., Kugler, J. W., Bartel, J., Law, M., Bateman, M., Klatt, N. E. et al. (1994). Prospective evaluation of prognostic variables from patient-completed questionnaires. {\sl Journal of Clinical Oncology} 12: 601--607.
\bibitem[Marchenko and Genton(2010)]{MG10} Marchenko, Y. V. and Genton, M. G. (2010). Multivariate log-skew-elliptical distributions with applications to precipitation data. {\sl Environmetrics} 21: 318-40.
\bibitem[Marshall and Olkin(1997)]{MO97} Marshall A. W. and Olkin I. (1997). A new method for adding a parameter to a family of distributions with application to the exponential and Weibull families. {\sl Biometrika} 84: 641--652.
\bibitem[Meeker and Escobar(1998)]{ME98} Meeker, W. Q. and Escobar, L. A. (1998). {\sl Statistical Methods for Reliability Data}. Wiley, New York.
\bibitem[Mitra(2012)]{M12} Mitra, D.(2012). Likelihood inference for left truncated and right truncated censored lifetime data. PhD Thesis, McMaster University.
\bibitem[Mitzenmacher(2001)]{M01} Mitzenmacher, M. (2001). A brief history of generative models for power law and lognormal distributions. {\sl Internet Mathematics} 1: 226--251.
\bibitem[Mudholkar and Hutson(2000)]{MH00} Mudholkar, G. S. and Hutson, A. D. (2000). The epsilon-skew-normal distribution for analyzing near-normal data. {\sl Journal of Statistical Planning and Inference} 83: 291--309.
\bibitem[Nadarajah and Bakar(2013)]{NB13} Nadarajah, S. and Bakar, S. A. A. (2013). CompLognormal: An R Package for Composite Lognormal Distributions. {\sl A peer-reviewed, open-access publication of the R Foundation for Statistical Computing} 97.
\bibitem[R Core Team(2013)]{R13} R Core Team (2013). {\sl R: A language and environment for statistical computing}. R Foundation for Statistical Computing, Vienna, Austria.  URL \url{http://www.R-project.org/}.
\bibitem[Roy and Dey(2015)]{RD15} Roy, V., and Dey, D. K. (2015). Propriety of posterior distributions arising in categorical
and survival models under generalized extreme value distribution. {\sl Statistica Sinica} 24: 699--722.
\bibitem[Rubio et al.(2015)]{R15} Rubio, F. J., Ogundimu, E. O., and Hutton, J. L. (2015). On modelling asymmetric data using two–-piece sinh--arcsinh distributions. {\sl Brazilian Journal of Probability and Statistics}, in press.
\bibitem[Rubio and Steel(2014)]{RS14} Rubio, F. J. and Steel, M. F. J. (2014). Inference in Two-Piece Location-Scale models with Jeffreys Priors (with discussion). {\sl Bayesian Analysis} 9: 1--22.
\bibitem[Vallejos and Steel(2015)]{VS15} Vallejos, C. and Steel, M. F. J. (2015). Objective Bayesian Survival Analysis Using Shape Mixtures of Log-Normal Distributions. {\sl Journal of the American Statistical Association} 110: 697--710.
\end{thebibliography}
\end{document}